\documentclass[a4paper]{article}
\usepackage[T1]{fontenc} 
\usepackage{float}       
\usepackage{helvet}      
\usepackage{mathptmx}    
\usepackage{setspace}    

\usepackage{amsmath}
\usepackage{amssymb}
\usepackage{epsfig}
\usepackage{graphicx}
\usepackage{dcolumn}
\usepackage{bm}
\newlength{\colsize}
\newcommand{\Tg}{\ensuremath{T_{\rm g}}}
\newcommand{\Tc}{\ensuremath{T_{\rm c}}}

\newcommand{\Rg}{\ensuremath{R_\mathrm{g}}}
\newcommand{\Ree}{\ensuremath{R_\mathrm{e}}}

\newcommand{\film}{\ensuremath{h}}

\newcommand{\epsilonmech}{\varepsilon}
\newcommand{\epseng}[1][]{\ensuremath{{\epsilonmech}_{\mathrm{eng}#1}}} 
\newcommand{\sy}{\ensuremath{\sigma_Y}} 

\newcommand{\fref}[1]{Fig.~\ref{#1}}
\newcommand{\Fref}[1]{Figure~\ref{#1}}
\newcommand{\eref}[1]{Eq.~\ref{#1}}

\title{Molecular dynamics simulations of glassy polymers}

\author{%
Jean-Louis Barrat\thanks{%
Universit\'e de Lyon; Univ. Lyon I,  Laboratoire de
Physique de la Mati\`ere Condens\'ee et des Nanostructures; CNRS,
UMR 5586, 43 Bvd. du 11 Nov. 1918, 69622 Villeurbanne Cedex,
France%
}%
\and
J\"org Baschnagel\thanks{
Institut Charles Sadron, Universit\'e de Strasbourg, CNRS UPR 22, 23 Rue du Loess, BP 84047, 67034 Strasbourg Cedex 2, France}
\and
Alexey Lyulin
\thanks{Group Theory of Polymers and Soft Matter, Eindhoven Polymer Laboratories,
Technische Universiteit Eindhoven, P.O. Box 513 5600 MB Eindhoven, the Netherlands,
and   Dutch   Polymer   Institute,   P.O.   Box   902,   5600   AX   Eindhoven,   The
Netherlands. }}

\begin{document}
\maketitle
\begin{abstract}
We review recent results from computer simulation studies of polymer glasses, from  chain dynamics around glass transition temperature \Tg \  to the  mechanical behaviour below \Tg.  These results clearly show that  modern computer simulations are able to address and give clear answers to some important issues in the field, in spite of the obvious limitations in terms of length and time scales. In the present review we discuss the cooling rate effects, and dynamic slowing down of different relaxation processes when approaching \Tg \  for both model and chemistry-specific polymer glasses. The impact of geometric confinement on the glass transition is discussed in detail. We also show that computer simulations are very useful tools to study structure and mechanical response of glassy polymers. The influence of large deformations on mechanical behaviour of polymer glasses in general, and strain hardening effect in particular are reviewed. Finally, we suggest some directions for future research, which we believe will be soon within the capabilities of state of the art computer simulations, and correspond to problems of fundamental interest.
\end{abstract}


\section{Introduction}
A polymer is a macromolecular chain resulting from the connection of a large number of monomeric units. The number of monomers (the chain length $N$) typically ranges between $10^3$ and $10^5$ in experimental studies of polymer melts \cite{LodgeMuthu_JPC1996}.  Recently, this range of chain lengths has also become accessible in simulations \cite{Theodorou:CES2007,WittmerEtal:PRE2007}.  Such a long chain has an open structure which is strongly pervaded by other chains in the melt \cite{RubinsteinColby}. The strong interpenetration of the chains has important consequences for the properties of the melt.  For instance, intrachain excluded volume interactions, which swell the polymer in dilute solution \cite{LodgeMuthu_JPC1996,RubinsteinColby}, are almost screened so that a chain  behaves on large length scales approximately as a random coil \cite{WittmerEtal:PRE2007}.  Furthermore, chain interpenetration also impacts the polymer dynamics by creating a temporary network of entanglements. Entanglements strongly slow down the chain relaxation and make the melt viscoelastic already at high temperature \cite{McLeish_AdvPhys2002,Likhtman:JNNFM2009}.

\begin{figure}
\includegraphics*[width=\colsize]{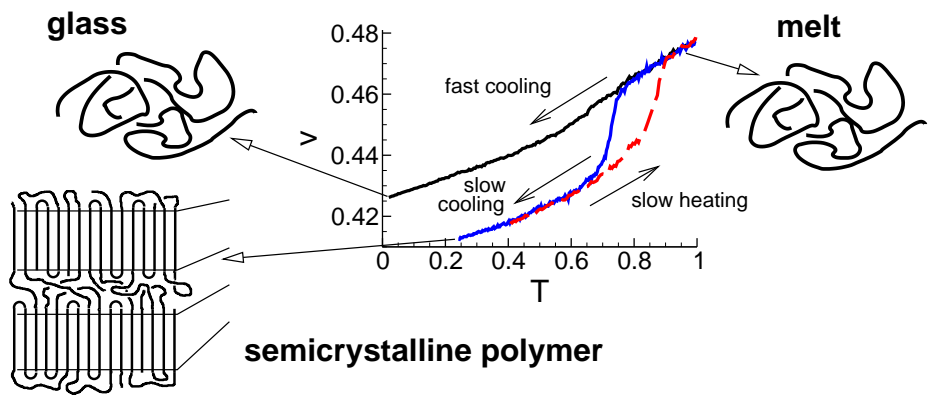}
\caption[]{Volume-temperature diagram of a coarse-grained model for poly(vinyl alcohol) \cite{MeyerMuellerPlathe:JCP2001,MeyerMuellerPlathe:Macromolecules2002}. The volume per monomer ($v$) and the temperature ($T$) are given in Lennard-Jones units. In the liquid phase the chains have random-coil-like configurations and the structure of the melt is amorphous.  The amorphous structure is preserved on rapid cooling and the melt eventually undergoes a glass transition at low $T$.  For slow cooling the melt transforms into a semicrystalline material in which sections of folded chains order in lamellar sheets that coexist with amorphous regions.  When the crystal is slowly heated up (dashed line), it melts at a higher $T$ than the temperature where crystallization occurs. This hysteresis is typical of first-order phase transitions.}
\label{fig:polycyc}
\end{figure}

On cooling toward low temperature the polymer melt eventually transforms into a solid. Polymeric solids are either semicrystalline or glassy (\fref{fig:polycyc}) \cite{StroblPolymerPhysics}.  In the semicrystalline state amorphous regions are intercalated between crystalline lamellar sheets.  The sheets consist of chains which are folded back on themselves so that chain sections align parallel to each other \cite{StroblPolymerPhysics,Muthukumar_AdvChem2004}. The ability to form crystals crucially depends on the polymer microstructure.  Only chains with regular configurations, e.g.\ isotatic or syndiotatic orientations of the sidegroups or chains without sidegroups, can fold into crystalline lamellae.  However, even in these favorable cases full crystallization is hard to achieve \cite{Muthukumar_AdvChem2004}.

This intrinsic difficulty to crystallize favors glass formation \cite{mckenna,AngellEtal_JAP2000,Donth}. Polymer melts either can be easily supercooled (\fref{fig:polycyc}) or, due to the irregular chain structure, do not crystallize at all.  An example for the former case is bisphenol-A polycarbonate (PC) \cite{Sundararajan1985}, while examples for the latter involve homopolymers with atactic (bulky) sidegroups, such as atactic polystyrene (PS), or random copolymers, such as {\em cis-trans}-1,4-polybutadiene (PBD). These polymeric glass formers share with other (intermediate and fragile) glass-forming liquids the key characteristic feature of the glass transition; that is, little change of the amorphous structure, but a huge non-Arrhenius-like slowing down of the dynamics on cooling toward the glass transition temperature $\Tg$ \cite{AngellEtal_JAP2000,Donth,kob-book}. Understanding the molecular origin of this disproportionate behaviour represents a great scientific challenge \cite{AngellEtal_JAP2000,Donth,kob-book,DebenedettiStillinger:Nature2001,GoetzeBook2009}. In addition to this fundamental interest in the study of the glass transition, solid polymers are also integral components in many modern applications \cite{Fredrickson:Nature2008}. For instance, glassy polymeric materials are appreciated because of their spectacular mechanical properties \cite{Ferry,Haward97,WardHadley_Book}. Instead of failing abruptly when subject to strong deformations, some polymers, such as polycarbonate, may harden for large strains, leading to a tough mechanical response. A microscopic understanding, elucidating the structure-property relationship, of this behaviour is still elusive.  A further example is provided by thin polymer films, extensively used in technological applications as protective coatings, optical coatings, adhesives, etc. In the ongoing quest for progressively smaller structures and devices these films may attain nanoscopic dimensions, where deviations from the bulk behaviour should be expected \cite{GranickEtal:JPSB}. Indeed, many recent studies suggest that the $\Tg$ of thin films is shifted by spatial confinement, a striking observation which is not well understood yet \cite{ForrestDalnoki:AdvColSci2001,AlcoutlabiMcKenna:JPCM2005}.

This wide array of challenging questions---from the glass transition to the impact of external stimuli---bestows the theoretical understanding of glassy polymers with particular significance. Molecular simulations can contribute to this research. Over the past two decades, computational models and methods have been developed for simulating these glassy systems. The progress made in the field has regularly been the subject of topical reviews. For instance, detailed reports of chemically realistic modeling approaches may be found in Refs.~\cite{Clarke_review1995,PaulSmith_RPP2004,RottlerJPCM2009}, while work on coarse-grained models is reviewed in Refs.~\cite{binder02,binder02b,binder03,binder04,baschnagel05,RottlerJPCM2009}. The purpose of the present article is to give a brief account of recent work.

We begin our survey by a short introduction to the modeling of polymers in simulations (section~\ref{sec:model}). The glass transition temperature naturally splits the following discussion into two parts, a part devoted to the properties of the model glass formers above $\Tg$ (section~\ref{sec:aboveTg}) and a part addressing sub-$\Tg$ phenomena (section~\ref{sec:belowTg}). In section~\ref{sec:aboveTg}, our discussion is mainly concerned with the dynamics of weakly supercooled polymer melts and the impact of spatial confinement on their behaviour. Section~\ref{sec:belowTg} focuses on the response of polymeric glasses to both weak and large external deformation, where respectively the linear mechanical behaviour of the glassy melt and its approach to material failure will be explored. The article concludes with a short summary of the presented results and an outlook on possible future research directions.

\section{Computer simulations: models and computational aspects}
\label{sec:model}

Molecular simulations of glass-forming polymers face the important challenge that the structure and dynamics of these systems are governed by a large spread of length and time scales \cite{Binder_MCMD1995,NielabaEtal:Springer2002}. The relevant length scales extend from the atomic diameter ($\sim 10^{-10}$ m) to the chain dimension ($\sim 10^{-7}$ m for $N \sim 10^4$). The spread of time scales is even larger; it ranges from bond vibrations ($\sim 10^{-13}$ s) to the slow structural relaxation close to $\Tg$ ($\sim 10^2$ s).

From the inspection of these scales, it is clear that simulation approaches have some limitations and also require simplifications.  An obvious limitation are the accessible time scales.  With the currently available computer power the longest time that can be reached in molecular dynamics (MD) simulations is roughly a few $\mu$s.  This time scale is 8 orders of magnitude smaller than the relaxation time at the experimental $\Tg$, implying that the simulated $\Tg$ is shifted by about 25 degrees to higher temperature ($T$) relative to experiments (where we used the rule of thumb: 1 decade in time $\widehat{\approx}$ 3 K). The simplifications that are currently necessary concern the simulation models which are obtained by some kind of coarse-graining procedure, designed to eliminate (some) fast degrees of freedom by incorporating them in effective potentials. This model-building step can take different levels of complexity \cite{JBETal:AdvPolySci2000,MullerPlathe:2002,Theodorou:CES2007,PeterKremer:SoftMatter2009}, but roughly speaking, there are two families, atomistic models and coarse-grained models. In the following, we briefly introduce these models and discuss their strengths and weaknesses as we go.

Before doing so, however, we want to point out some general appealing features of the simulations to balance the reservations expressed above. The simulations offer full control over the perfectly defined system under study. For instance, it is possible to vary in a systematic manner parameters, such as chain stiffness, chain length, etc., while keeping all other defining properties of the model, so as to single out the impact of each parameter on the properties of the system. Furthermore, the simulations enable one to study local properties or to explore correlations which are hard (or even impossible) to access in experiments. This provides microscopic insight and a means to test theoretical concepts. Examples from the recent literature involve investigations of the potential energy landscape of glass-forming materials \cite{tsalikis08,tsalikis08a,Heuer:JPCM2008}, of single-molecule diffusion in polymeric matrices \cite{vallee07,vallee06} or of fracture in glassy polymers \cite{RottlerJPCM2009}

\paragraph{Atomistic models.}  Atomistic models replace electronic degrees of freedom by force fields, i.e., empirical potentials for bond-length, bond-angle, torsional-angle, and nonbonded interactions, whose parameters are determined from quantum-chemical calculations and experiments \cite{PaulSmith_RPP2004,JBETal:AdvPolySci2000}. During the past decades, force fields have been developed for both explicit atom (EA) models, treating every atom present in nature as a separate interaction site, and united atom (UA) models which represent a small number of real atoms (e.g., $\mathrm{CH}_2$ or $\mathrm{CH}_3$) by a single site.  The reduction of the number of interaction sites in the UA model has the computational advantage of allowing longer simulation times.  With a time step of $\sim 10^{-15}$ s a few thousand united atoms can be simulated over several 100 ns, about an order of magnitude longer than an EA simulation of comparable system size.

Both EA and UA models were employed to study the properties of glassy polymers (see e.g.\ \cite{Clarke_review1995,PaulSmith_RPP2004,RottlerJPCM2009} for reviews).  Recent examples include polyisoprene (EA model \cite{ColmeneroEtal:PRE2002,ColmeneroEtal:JPCM2003}), atactic PS (UA model \cite{lyulin02,lyulin02b,lyulin03,Lyulin2005,lyulin07}), bisphenol-A PC (UA model \cite{Lyulin2005,lyulin07}) 1,4-PBD (UA and EA models \cite{PaulSmith_RPP2004,SmithEtal:JCP2004,KrushevPaul_PRE2003,KrushevEtal_Macromolecules2002,PaulEtal:PRE2006,SmithBedrov:JPSB2007,ColmeneroEtal:JPCM2007,NarrosEtal:JCP2008}), and poly(ethylene oxide) or atactic poly(propylene oxide) (EA models \cite{Vogel:Macro2008}). Certainly, the main strength of these modeling efforts is that the simulation results allow for a direct comparison with experiments. In some cases, such a comparison is possible with commercial software packages \cite{NarrosEtal:JCP2008}, while a careful fine-tuning of the force field is often required \cite{SmithEtal:JCP2004,SmithBedrov:JPSB2007}. Experience from those modeling approaches---not only for polymers but also e.g.\ for amorphous $\mathrm{SiO}_2$ \cite{binder04,kobreview1999}---suggests that the design of a chemically realistic model, aiming at a parameter-free comparison between simulation and experiment, should involve information about both structural and dynamic properties \cite{PaulSmith_RPP2004}.

\paragraph{Generic models.}  Atomistic simulations are ideally suited to study specific polymers, including their glass transition and the properties of the glassy state.  On the other hand, the strong increase of the relaxation time, which eventually leads to vitrification on cooling through $\Tg$, is common to all glass-forming polymers, irrespective of their chemical composition and architecture.  This universal aspect suggests to employ simplified simulation models which only retain generic features of a polymer. Various such generic models---on a lattice \cite{binder04,binder03,binder02,binder02} or in the spatial continuum \cite{binder04,binder03,baschnagel05}---have been studied for glass-forming polymers. In the continuum, the simplest model are highly flexible bead-spring models, where spherical monomers are tethered together by springs and have nonbonded, Lennard-Jones (LJ) interactions \cite{binder04,binder03,baschnagel05}. More chemical realism can be introduced by making the chains semiflexible, through the addition of a bond-angle potential \cite{DotsonEtal:JCP2008} and, possibly also, of a torsional potential \cite{BulacuGiessen:PRE2007,BernabeiEtal:PRL2008,BernabeiEtal:2009}. Due to their simplicity, these generic models allow for an efficient simulation. This is important if one wants to vary the cooling rate \cite{buchholz02} or the strain rate \cite{RottlerJPCM2009} over decades, to explore systematically the impact of model parameters, such as chain length or chain flexibility \cite{DotsonEtal:JCP2008,BulacuGiessen:PRE2007,BernabeiEtal:PRL2008,BernabeiEtal:2009}, or to obtain good statistics for comparison with theory \cite{chong07}. Computational expedience is also important for exploratory studies of more complex systems, such as inhomogeneous polymer systems (e.g., polymer films, nanocomposites, semi-crystalline polymers) or glassy polymer mixtures (e.g., dynamically asymmetric polymer-polymer mixtures or polymer-solvent systems).  Some examples will be discussed in Sects.~\ref{sec:aboveTg} and \ref{sec:belowTg}.

\paragraph{Hierarchical models.}  For generic models, the gain in the accessible length and time scales is obtained at the expense of a loss of correlation to the atomistic conformation of the polymer.  For many problems in materials research, this loss is undesirable.  Therefore, much research efforts currently goes into the development of hierarchical approaches consisting of interconnected levels of modeling (atomistic, generic, macroscopic) \cite{Theodorou:CES2007,PeterKremer:SoftMatter2009,McCartyEtal:JPCB2009,MurtolaEtal:PCCP2009}. The idea is that each level treats phenomena on its specific length and time scales and then passes on the results as input to the next, more coarse-grained level, until the desired materials properties can be predicted. Such multiscale simulation methods represent a powerful approach whose potential for the modeling of glassy polymers is beginning to be explored (see e.g.\ \cite{StrauchEtal:PCCP2009}).

\paragraph{Remarks on simulation methods.} Sections~\ref{sec:aboveTg} and \ref{sec:belowTg} will present results from molecular dynamics (MD) simulations, a numerical method to integrate the classical equations of motion for a many-body system in a given thermodynamic ensemble \cite{FrenkelSmit,binder04b}. Therefore, MD is the natural simulation technique to address dynamical problems, such as the glass transition. However, the realistic MD dynamics carries an obvious price: the equilibration time for the system under consideration---for instance, for a long-chain polymer melt close to its $\Tg$---will exceed the maximum time of a few microseconds one is currently able to simulate.

Here Monte Carlo (MC) techniques may provide a promising avenue because of large freedom to design MC moves \cite{FrenkelSmit,LandauBinder}. The hope is to find an efficient algorithm allowing one to decorrelate the configurations of glassy polymer melts rapidly  For long-chain polymer melts, this demand on the algorithm implies, already at high $T$, that the MC move should be nonlocal, i.e., it should modify the chain conformation at large scales, and it should not require empty space because the melt is a dense liquid. A promising algorithm satisfying these requirements employs double-bridging moves which alter the connectivity between two neighbouring chains while preserving the monodispersity of the chains (for recent reviews see e.g.\ \cite{WittmerEtal:PRE2007,Theodorou:CES2007}).  Such a connectivity-altering move drastically changes the conformation of the two chains involved and thus relaxes the length scales on the order of the chain dimension efficiently. However, it does not alter the local packing of the monomers. An inherent hazard of the algorithm therefore is that, if the move is attempted repeatedly, a successful double-bridging event is likely to annihilate one of its predecessors by performing the transition between two chains in the reverse direction.  To avoid this inefficiency the nonlocal chain updating should be complemented by a move which efficiently mixes up the local structure of the melt.  At low $T$, efficient relaxation of the liquid structure calls for a method which alleviates the glassy slowing down in general.  Thus, any algorithm achieving this aim in nonpolymeric liquids should also accelerate the equilibration of glassy polymer melts, provided that it can be generalized to respect chain connectivity.  At present, no technique has been established to solve this problem.  However, possible candidates could be parallel tempering \cite{BunkerDunweg:PRE2000,YamamotoKob2000,DeMicheleSciortiono:PRE2002}, Wang-Landau sampling \cite{LandauBinder} or variants thereof \cite{YanEtal:PRL2004}, or transition path sampling methods \cite{ChopraEtal:JCP2008}.

\section{Glass transition and properties of the supercooled polymer liquid}
\label{sec:aboveTg}

\subsection{Bulk properties}

\subsubsection{Glass transition temperature}
\label{subsubsec:Tg}
In polymer melts, the transition from the glass to the liquid is accompanied by a strong decrease in the shear modulus, typically of three to four orders of magnitude \cite{WardHadley_Book}. It is thus clear that the glass transition temperature is a characteristic of high engineering relevance. Classical methods for the experimental determination of $\Tg$ are calorimetry and dilatometry. Both methods hint at the kinetic features of the glass transition. The transition occurs when the relaxation time for volume recovery (dilatometry) or enthalpy recovery (calorimetry) becomes longer than the time scale of the experiment (i.e., than the cooling or heating rate) \cite{mckenna,AngellEtal_JAP2000,Donth}. Upon cooling the polymer liquid falls out of equilibrium close to $\Tg$ and freezes below $\Tg$ in a glassy state, the properties of which depend on the details of the cooling process and tend to age physically during further isothermal equilibration \cite{mckenna,AngellEtal_JAP2000,Donth}.  The glass transition temperature is located in the $T$ interval where the polymer melt smoothly evolves from the liquid to the solid state, and can be defined operationally through some prescription, for instance, as the intersection of straight-line extrapolations from the glassy and liquid branches of the volume-temperature curve \cite{mckenna,Donth}.

\begin{figure}
\includegraphics*[width=\colsize]{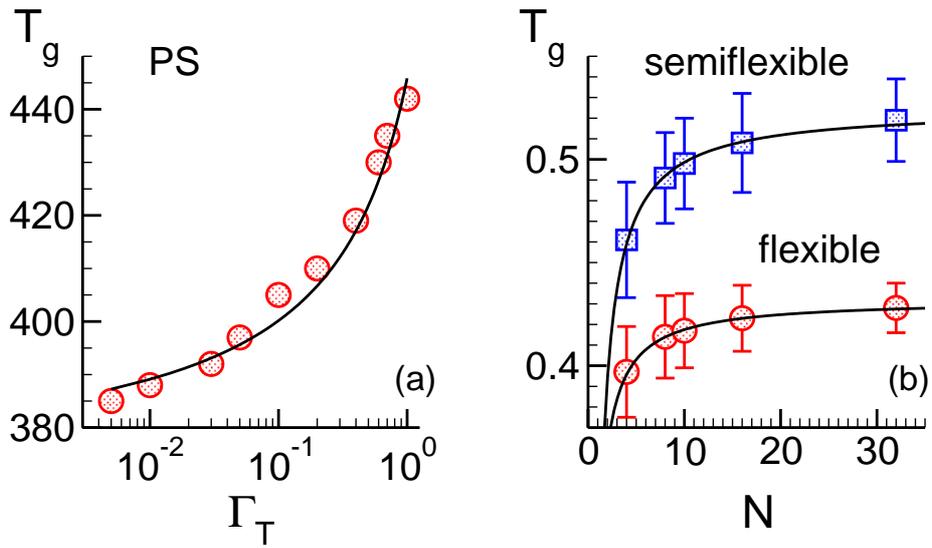}
\caption[]{Panel (a): Glass transition temperature ($\Tg$) versus cooling rate ($\Gamma_T$) for an atomistic model of atactic PS \cite{lyulin03}. The simulation box contains 8 chains with $N=80$ monomers each. The solid line shows a fit to \eref{eq:TggammaT} with $T_0=371$ K, $B=110$ K, and $A=0.23$ ps/K.  Panel (b): $\Tg$ versus chain length ($N$) for a fully flexible and a semiflexible (with angular potential) bead-spring model \cite{SchnellPhD}.  The simulation box contains at least 192 chains and the cooling rate is $\Gamma_T = 2 \times 10^{-5}$. The solid lines are results of a fit to \eref{eq:TgvsN} with $\Tg^{\infty}=0.432$, $K=0.145$ (flexible model) and $\Tg^{\infty}=0.525$, $K=0.264$ (semiflexible model). All data are given in Lennard-Jones units for the bead-spring models.}
\label{fig:Tg_vs_N_and_Gq}
\end{figure}

Similar extrapolation procedures are also applied in simulation studies \cite{PaulSmith_RPP2004,buchholz02,binder03,lyulin03,Lyulin2005,SolderaMetatla:PRE2006,MetatlaSoldera:Macro2007,SchnellPhD,VollmayrEtal:JCP1996,VollmayrEtal:PRB1996}. The resulting $\Tg$ values have dependences comparable to experimental ones despite the much larger cooling rate employed---typically $10^{12} \, \mathrm{K/min}$ in simulations and $10 \, \mathrm{K/min}$ in experiments. For instance, Soldera and Metatla find a linear relationship between numerical and experimental $\Tg$ values from atomistic simulations of various vinylic polymers \cite{SolderaMetatla:PRE2006}. \Fref{fig:Tg_vs_N_and_Gq} reveals that $\Tg$ decreases nonlinearly with the logarithm of the cooling rate ($\Gamma_T$) \cite{PaulSmith_RPP2004,buchholz02,binder03,lyulin03},
\begin{equation}
 \Tg (\Gamma_T) = \Tg^0 - \frac{B}{\ln (A \Gamma_T)} \;,
\label{eq:TggammaT}
\end{equation}
in accordance with experimental observation \cite{BrueningSamwer:PRB1992}.  Also in agreement with experiment \cite{mckenna,DingEtal:Macromolecules2004,KunalEtal:Macromolecules2008,HintermeyerEtal:Macromolecules2008,AgapovSokolov:Macromolecules2009}, $\Tg$ increases with chain rigidity and chain length \cite{BulacuGiessen:PRE2007,lyulin03,lobe94}. The chain length dependence can be fitted to the empirical Fox-Flory equation
\begin{equation}
\Tg(N) = \Tg^{\infty} - \frac{K}{N} \;,
\label{eq:TgvsN}
\end{equation}
which usually describes experimental data well (if the molecular weight is not too small \cite{mckenna}), although other forms have recently been discussed in the literature \cite{HintermeyerEtal:Macromolecules2008,AgapovSokolov:Macromolecules2009} and can be rationalized theoretically \cite{DudowiczEtal:ACP2008}.

\subsubsection{Dynamics of the supercooled melt}
\label{subsubsec:dynamics}

\begin{figure}
\includegraphics*[width=\colsize]{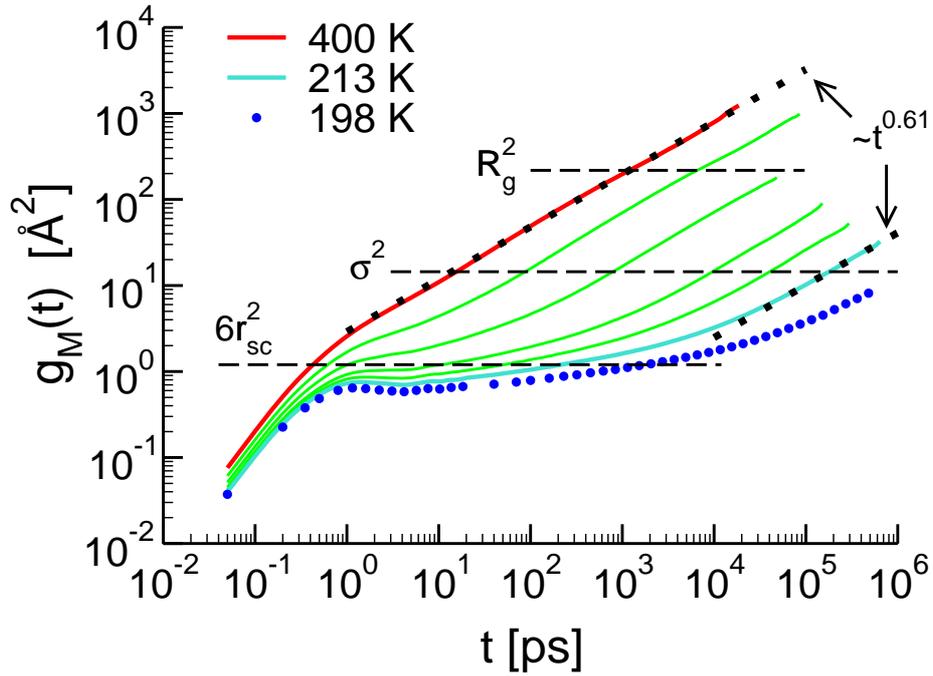}
\caption[]{Mean-square displacement (MSD) $g_\mathrm{M}(t)$ versus time $t$ for a united atom model of 1,4-PBD (results adapted from \cite{PaulEtal:PRE2006}, with permission). The MSD is averaged over all united atoms in the melt containing 40 chains of 30 repeats units. The temperatures shown are (from left to right): $T=400$ K, 323 K, 273 K, 240 K, 225 K, 213 K, and 198 K ($\Tc \approx 214$ K). The horizontal dashed lines indicate the radius of gyration $\Rg^2 \approx 218$ {\AA}$^2$, the average Lennard-Jones diameter of the united atoms $\sigma \approx 3.8$ {\AA}, and an estimate for the Lindemann localization length $r_\mathrm{sc} \approx 0.45$ {\AA} \cite{ColmeneroEtal:PRE2002}. The dotted lines for $T=400$ K and 213 K show the power law $\sim t^{0.61}$, characteristic of Rouse-like motion.}
\label{fig:msd_PB}
\end{figure}

The precursor of the glass transition is the strong slowing down of structural relaxation processes on approach to the transition from the liquid. This dynamical feature is a hallmark of strongly interacting disordered matter, including dense colloids \cite{Schweizer:Current2007}, granular materials \cite{KeysEtal:NatPhys2007}, and supercooled liquids \cite{AngellEtal_JAP2000,DebenedettiStillinger:Nature2001}. In simulated polymer melts, it is observed in all dynamic correlation functions, e.g., in dynamic structure factors, conformational correlation functions, or dielectric relaxation \cite{PaulSmith_RPP2004}.

\begin{figure}
\includegraphics*[width=\colsize]{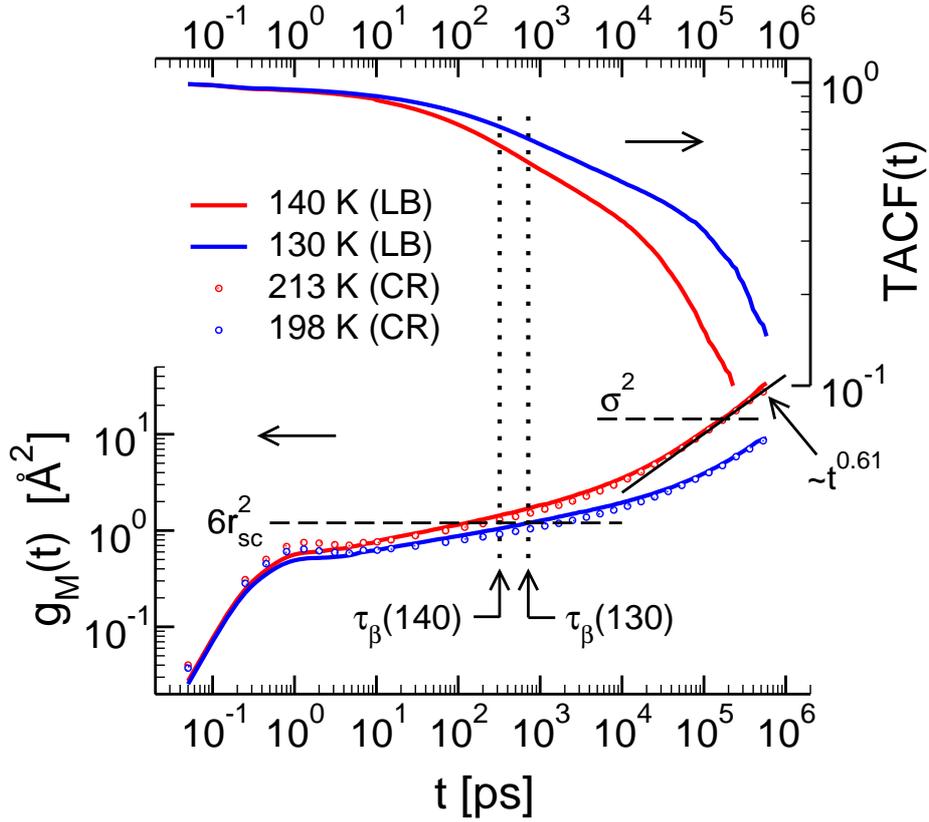}
\caption[]{Monomer MSD $g_\mathrm{M}(t)$ and torsional autocorrelation function (TACF) versus $t$ for 1,4-PBD. The data for $T=130$ K and 140 K (lines) are obtained from a model with reduced torsional barriers (LB), but which is otherwise the same as in \fref{fig:msd_PB} (results adapted from \cite{SmithBedrov:JPSB2007}, with permission). For comparison the symbols show the MSDs at $T=198$ K and 213 K for the chemically realistic (CR) model with the full torsional potential.  These temperatures are approximately at the same distance to $\Tg$ ($\approx 170$ K) as for the LB model ($\Tg \approx 102$ K) \cite{SmithBedrov:JPSB2007}. The horizontal dashed lines indicate the average Lennard-Jones diameter of the united atoms $\sigma \approx 3.8$ {\AA} and an estimate for the Lindemann localization length $r_\mathrm{sc} \approx 0.45$ {\AA} \cite{ColmeneroEtal:PRE2002}. The solid line for $T=140$ K shows the power law $\sim t^{0.61}$, characteristic of Rouse-like motion. The vertical dotted lines represent the $\beta$-relaxation times ($\tau_\beta =716.5$ ps for $T=130$ K, $\tau_\beta =319.4$ ps for $T=140$ K).}
\label{fig:msd+tacf_ReducedBarriers_PB}
\end{figure}

As an example, \fref{fig:msd_PB} shows the mean-square displacement (MSD) $g_\mathrm{M}(t)$, averaged over all monomers of a chain, for a chemically realistic model of 1,4-PBD \cite{PaulEtal:PRE2006}. At high temperature, the MSD directly crosses over from ballistic motion ($\sim t^2$) at short times to subdiffusive motion ($\sim t^{x_0}$ with $x_0 \approx 0.61$) at intermediate times where the MSD is bound between the monomer size ($\sigma$) and the end-to-end distance $\Ree$ of a chain. This subdiffusive motion does not depend on the strength of the torsional barrier (cf.\ \fref{fig:msd+tacf_ReducedBarriers_PB}), is present even if the torsional potential is absent \cite{KrushevEtal_Macromolecules2002}, and thus reflects the universal Rouse-like dynamics of nonentangled chains in a polymer melt \cite{RubinsteinColby}. The term ``Rouse-like'' stresses the fact that simulations of nonentangled chains \cite{PaulSmith_RPP2004,chong07,bennemann99a,BrodeckEtal:JCP2009} find both accord with Rouse predictions---e.g., the Rouse modes are (nearly) orthogonal for all $t$---and deviations from them---e.g., the Rouse modes are stretched with stretching exponents depending on the mode index and the MSD of the chain's center of mass increases sublinearly for intermediate times. The origin of these deviations is not fully understood.  However, for short chains the relaxation time of a chain is not well separated from the (local) $\alpha$-relaxation so that finite-$N$ corrections to the Rouse behaviour must be expected \cite{chong07}.  Moreover, even for long chains intermolecular interactions between the polymers are not completely screened on mesoscopic length scales, which may cause subdiffusive center-of-mass motion \cite{ZamponiEtal:JPCB2008} or deviations from reptation theory \cite{Likhtman:JNNFM2009}, and lead to corrections to chain ideality \cite{WittmerEtal:PRE2007,MeyerEtal:EPJE2008}.

On cooling towards $\Tg$ the Rouse-like motion shifts to progressively longer times due to the appearance of a plateau regime. In this regime, the MSD increases only very slowly with time and is of the order of 10\% of the monomer diameter, reflecting the temporary localization of a monomer in the cold melt. For polymer models with intramolecular rotational barriers, such as PBD, this intermittence of large scale motion stems from two mechanisms of dynamical arrest \cite{PaulSmith_RPP2004,KrushevPaul_PRE2003,ColmeneroEtal:JPCM2007}: monomer caging by near neighbours in the dense melt and the slowing down of intrachain conformational transitions occurring through (correlated) torsional motion.

Recent work by Smith and Bedrov \cite{SmithBedrov:JPSB2007} suggests that the interplay of both mechanisms can explain the Johari-Goldstein $\beta$ relaxation in polymers \cite{Donth}. \Fref{fig:msd+tacf_ReducedBarriers_PB} reproduces one of their results. Since experiments reveal that the separation of the $\alpha$ and $\beta$ processes occurs only on time scales significantly longer than the multiple microsecond trajectories generated for the chemically realistic (CR) PBD model, Smith and Bedrov employ a PBD model with reduced torsional barriers (LB model) but otherwise identical interactions as for the CR model. While structural properties of the PBD melt remain unaffected \cite{KrushevEtal_Macromolecules2002,KrushevPaul_PRE2003}, the reduction of the barriers accelerates the dynamics and shifts the $\alpha$-$\beta$ bifurcation into the simulation time window. \Fref{fig:msd+tacf_ReducedBarriers_PB} compares $g_\mathrm{M}(t)$ with the decay of the torsional autocorrelation function (TACF). Similar to $g_\mathrm{M}(t)$, the TACF relaxes in two steps. To both steps can be associated relaxation times, $\tau_\beta$ and $\tau_\alpha$, which display an Arrhenius ($\tau_\beta$) and a non-Arrhenius ($\tau_\alpha$) increase with decreasing $T$, characteristic of $\beta$ and $\alpha$ processes, respectively. At low $T$, both processes are well separated from each other, the beta process corresponding to times for which the monomers displace, on average, by about 10\% of their diameter. This implies that the cages imposed by the polymer matrix remain largely intact during the $\beta$ process and create a potential energy landscape for the underlying conformational transitions. Indeed, detailed analysis reveals that (nearly) all dihedrals visit all torsional states during the $\beta$ relaxation. However, these visits do not occur with the equilibrium probability because the propensity of a dihedral to return to a preferred conformational state increases on cooling, due to the stiffening of the matrix. Equilibrium occupancy of conformational states is only achieved for times comparable to the $\alpha$ time scale when the ``cage effect'' of the matrix fully decays (see also \cite{NarrosEtal:JCP2008,Vogel:Macro2008} for similar results).

\begin{figure}
\includegraphics*[width=\colsize]{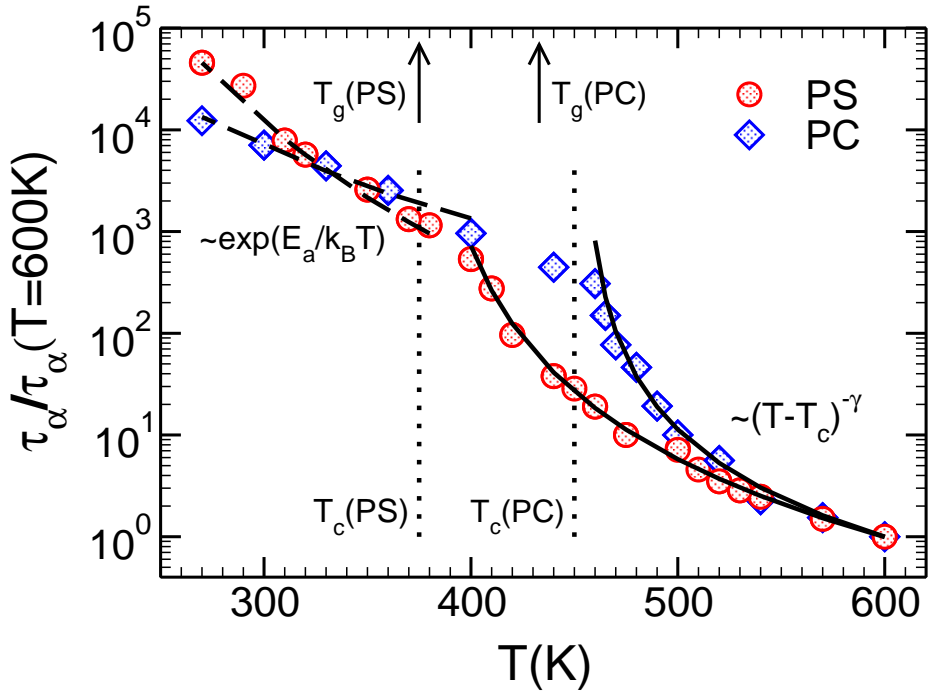}
\caption{Temperature dependence of the $\alpha$ relaxation time ($\tau_\alpha$) for united atom models of atactic PS and bisphenol-A PC. $\tau_\alpha$ is normalized by its value at $T=600$ K for both polymers (PS: $\tau(600\,\mathrm{K}) = 1.4$ ps, PC: $\tau(600\,\mathrm{K}) = 1.3$ ps) and is obtained from the monomer MSD in simulations of 1 chain with $N=80$ for PS and 64 chains with $N=10$ for PC. For both polymers, the increase of $\tau_\alpha$ at high $T$ is compatible with the MCT prediction $\tau_\alpha \sim (T-\Tc)^{-\gamma}$ (PS: $\Tc \simeq 375$ K, $\gamma \simeq 3$; PC: $\Tc \simeq 450$ K, $\gamma \simeq 2.2$). However, the divergence at $\Tc$ is avoided and a crossover to an Arrhenius behaviour (PS: $E_\mathrm{a} \simeq 30$ kJ/mol; PC: $E_\mathrm{a} \simeq 16.8$ kJ/mol) occurs close to $\Tg$ (derived from dilatometry. PS: $\Tg \simeq 375$ K; PC: $\Tg\simeq 433$ K). This crossover indicates the transition from the cooperative translational dynamics above $\Tg$ to activated ($\beta$ process like) hoping below $\Tg$. Results adapted from \cite{Lyulin2005}.}
\label{fig:tau_TcTg_PS_PC}
\end{figure}

\begin{figure}
\includegraphics*[width=\colsize]{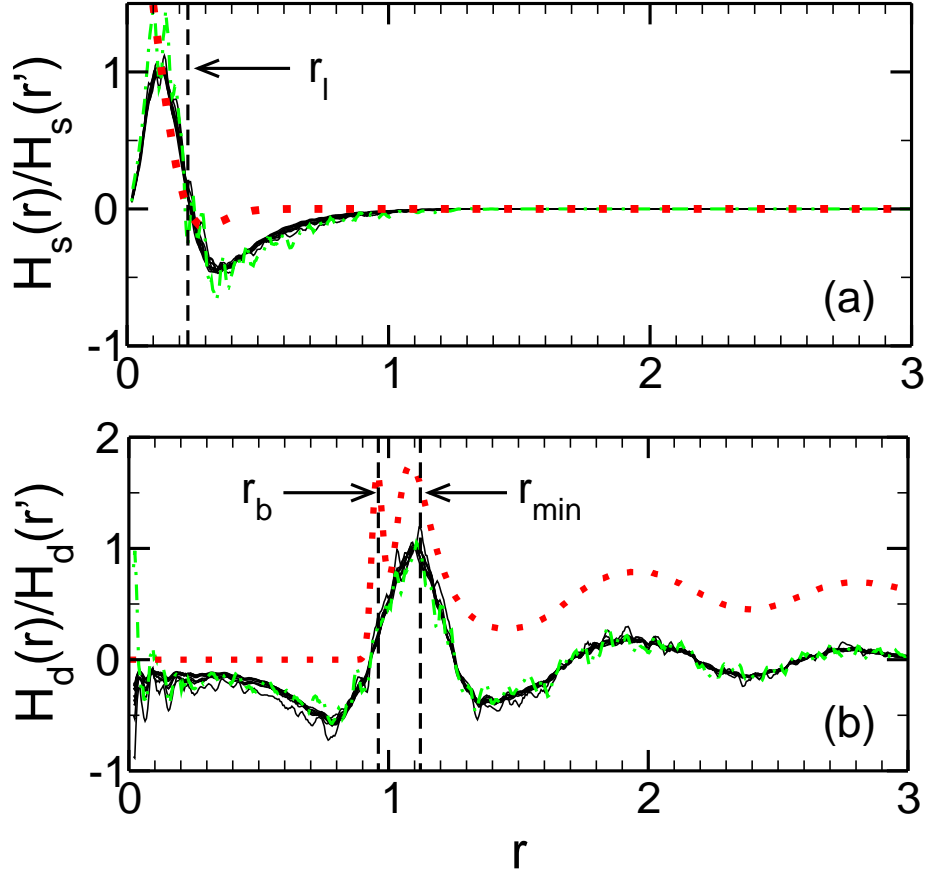}
\caption{Simulation results for a flexible bead-spring model: ${R}_\mathrm{x}(r,t)$ versus $r$ for different times from the plateau regime at $T=0.48$ ($\Tc \simeq 0.45$).  Panel (a) shows ${R}_\mathrm{x}(r,t)$ for the self-part of the van Hove function and panel (b) for its distinct part. In panel (b), the pair-distribution function  $g(r)$ is also shown (dotted line; rescaled to fit into the figure).  The first peak of $g(r)$ reflects the bond length ($\approx 0.97$) and the minimum position of the Lennard-Jones potential ($\approx 1.12$).  In panel (a), $r_\mathrm{l}$ ($=0.2323$) denotes the zero of ${R}_\mathrm{s}(r,t)$ and the circles represent a Gaussian approximation which has a zero at $\sqrt{6}r_\mathrm{sc} \simeq 0.2327$ and a minimum around $\sqrt{10}r_\mathrm{sc} \simeq 0.3$.  Here, $r_\mathrm{sc}$ ($\simeq 0.095$) is the Lindemann localization length.  In both panels, the dash-dotted lines correspond to the time closest to $t'$ which is the least precise because ${R}_\mathrm{x}(r,t)$ is undetermined for $t=t'$.  Adapted from \cite{aichele01b,baschnagel05}.}
\label{fig:Hr_self+dist_T048}
\end{figure}

In glass physics, the term ``cage effect'' is intimately connected to the mode-coupling theory (MCT) of the glass transition, which has certainly been one of the most influential theoretical approaches in the field during the last twenty years.  A comprehensive review of its foundations and applications was published recently \cite{GoetzeBook2009}.  By correlating the equilibrium structure to the dynamics of a glass-forming liquid MCT provides a framework for interpreting spatio-temporal correlations measured in experiment or simulation.  Analytical predictions are derived in the vicinity of an ideal glass transition which occurs at a critical temperature $\Tc$ and is driven by the mutual blocking of a particle and its neighbours (``cage effect'').  Extensive tests by experiments and simulations indicate that, although (the extrapolated) $\Tc$ lies above $\Tg$ and thus no structural arrest is observed at $\Tc$ (cf.\ \fref{fig:tau_TcTg_PS_PC}), MCT still describes many dynamical features well \cite{GoetzeBook2009,Goetze:JPCM1999}. Why this is so, represents a great challenge for the theoretical understanding \cite{Schweizer:Current2007,MayerEtal:PRL2006,KimKawasaki:JSM2008,Szamel:JCP2007,AndreanovEtal:EPL2009,Schweizer:Current2007,ChenEtal:JPCM2009}.

A key prediction of MCT is that structural relaxation functions should obey a factorization property in the plateau regime (also called $\beta$ regime in MCT). For the self ($G_\mathrm{s}$) and distinct ($G_\mathrm{d}$) parts of the van Hove correlation function \cite{HansenMcDonald} this reads
\begin{equation}
\label{eq:factorization_r}
G_\mathrm{x}(r,t) = F_\mathrm{x}(r) + H_\mathrm{x}(r) G(t) \quad
(\mbox{x = s, d}) \;.
\end{equation}
The factorization property refers to the fact that the correction to the nonergodicity parameter $F_\mathrm{x}(r)$ splits into two factors, of which $G(t)$ depends only on time (and temperature) and $H_\mathrm{x}(r)$ only on $r$ \cite{GoetzeBook2009,Goetze:JPCM1999}. Therefore, the ratio ($r' = \mbox{constant}$) \cite{SignoriniBarratKlein1990,KobAndersen_LJ_I_1995}
\begin{equation}
\label{eq:def_RX}
{R}_\mathrm{x}(r,t) = \frac{G_\mathrm{x}(r,t) - G_\mathrm{x}(r,t')}{G_\mathrm{x}(r',t) - G_\mathrm{x}(r',t')} = \frac{H_\mathrm{x}(r)}{H_\mathrm{x}(r')}
\end{equation}
should be independent of $t$. For a flexible bead-spring model \cite{aichele01b} \fref{fig:Hr_self+dist_T048} confirms this prediction directly from the simulation data (no fit) and additionally reveals the length scales pertinent for the dynamics in the plateau regime because distances for which $H_\mathrm{x}(r)$ is zero will not contribute to the relaxation. For the self part of the van Hove function the dynamics involves displacements up to the monomer diameter, whereas for the distinct part it includes monomers up to about the forth neighbour shell. This local character of the relaxation is a direct evidence for the cage effect.

The agreement between MCT and simulation, demonstrated in \fref{fig:Hr_self+dist_T048}, is not limited to the flexible bead-spring model. Simulations of semiflexible bead-spring models with rotational barriers \cite{BernabeiEtal:PRL2008,BernabeiEtal:2009} and of chemically realistic models, including 1,4 PBD \cite{PaulEtal:PRE2006,SmithEtal:JCP2004,ColmeneroEtal:JPCM2007}, PS and PC \cite{Lyulin2005}, generally find that many features of the spatiotemporal relaxation in weakly supercooled polymer melts (i.e., $T \gtrsim \Tc$) are well described by MCT (see however \cite{Vogel:Macro2008}). The main difference between these models and the flexible bead-spring model is that two coexisting arrest mechanisms---intramolecular barriers and monomer caging---determine the structural relaxation of the former, whereas only caging operates for the flexible model. This interpretation is a key result of Refs.~\cite{KrushevPaul_PRE2003,KrushevEtal_Macromolecules2002} and of the recent work by Bernabei {\em et al} \cite{BernabeiEtal:PRL2008,BernabeiEtal:2009} who also make the interesting conjecture that, within MCT, the pertinent theoretical framework for polymer models with internal rotational barriers are (so-called) higher-order transition scenarios \cite{GoetzeBook2009}, as it appears to be the case for other systems with distinct arrest mechanisms, such as dense colloidal suspensions with short-range attraction \cite{Sperl:PRE2003,ZacarelliPoon:PNAS2009} or polymer mixtures with strong dynamic asymmetry \cite{MorenoColmenero:JCP2006,MorenoColmenero:JPCM2007,ColmeneroArbe:SoftMatter2007}.

\begin{figure}
\includegraphics*[width=\colsize]{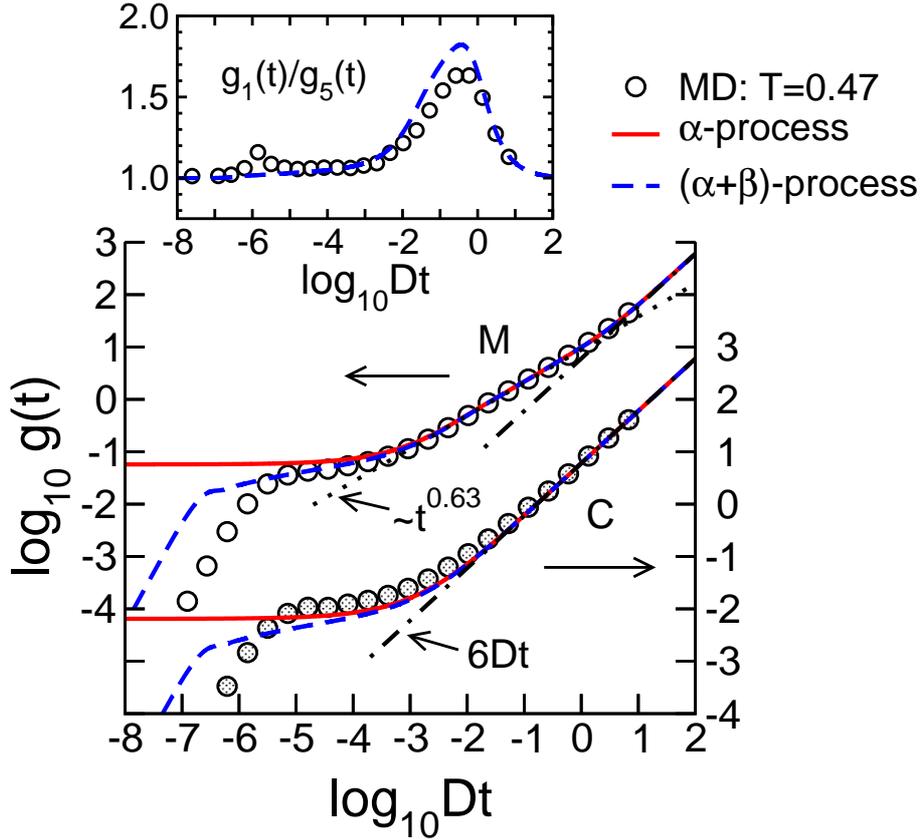}
\caption{Simulation results for a flexible bead-spring model: Log-log plot of the monomer MSD (labeled M, left scale) and the MSD of the chain's center of mass (labeled C, right scale) versus $Dt$ with $D$ being the diffusion coefficient of a chain. The inset shows the ratio $g_1(t)/g_5(t)$ (end monomer MSD over middle monomer MSD). The circle refer to the MD results at $T=0.47$ ($\Tc \simeq 0.45$), the solid lines to the MCT $\alpha$ master curve, and the dashed lines to the MCT predictions including the MCT $\beta$ process. The dash-dotted lines indicate the diffusive motion $6Dt$ and the dotted line shows the power law $\sim t^{0.63}$ of Rouse-like motion. Figure taken from \cite{chong07}.}
\label{fig:all_msd_MDT047_versus_MCT}
\end{figure}

A distinctive feature of MCT is that the dynamics (beyond the short time regime) is fully specified in terms of the liquid structure. This opens the way for an ``{\em ab initio}'' prediction of the simulated dynamics, solely based on static input obtained from an independent simulation of the studied glass former. Such an atomistic theory for the slow relaxation of (nonentangled) polymer melts has been developed \cite{ChongFuchs:PRL2002} and compared to simulations of a bead-spring model with $N=10$ \cite{aichele04,chong07}. The comparison gives semi-quantitative agreement between simulation and theory (for models with large rotational barriers further complications might arise \cite{BernabeiEtal:2009}). As an example, \fref{fig:all_msd_MDT047_versus_MCT} shows various MSDs, revealing the strengths and weaknesses of the approach \cite{chong07}. Certainly, the agreement for the monomer dynamics is very good. MCT describes the dynamics in the plateau regime, the following polymer-specific subdiffusive increase, $g_\mathrm{M} \sim t^{0.63}$ which is identified as a finite-$N$ deviation from Rouse behaviour, and the faster motion of the end monomer ($g_1$) relative to the central monomer ($g_5$) of the chain. On the other hand, the theory is not so satisfactory for the MSD of the chain's center of mass ($g_\mathrm{C}$). Besides underestimating the plateau height it does not reproduce the subdiffusive center-of-mass motion \cite{ZamponiEtal:JPCB2008} ($g_\mathrm{C} \sim t^{\approx 0.8}$) between the plateau and diffusive regimes.

The representation in \fref{fig:all_msd_MDT047_versus_MCT}---that is, plotting the data versus $Dt$ with $D$ being the chain's diffusion coefficient---facilitates the comparison of the time dependence in the $\beta$ and early $\alpha$-regimes, but camouflages a systematic deviation between theory and simulation. MCT predicts that the $\alpha$ relaxation time $\tau_\alpha(q)$ has the same $T$ dependence for all wave vectors $q$, whereas simulations, not only for flexible bead-spring models \cite{baschnagel05} but also for other glass formers \cite{FlennerSzamel:PRE2005_1,FlennerSzamel:PRE2005_2,FoffiEtal:PRE2004,VoigtmannEtal:2004,ChaudhuriEtal:PRL2007}, find that $\tau_\alpha$ for wave vectors smaller than the position $q^*$ of the first maximum of static structure factor increases on cooling more weakly than for $q \gtrsim q^*$.  This difference also implies a violation of the Stokes-Einstein relation \cite{MerabiaEtal:EPJE2002,KumarEtal:JCP2006,sokolov09}, i.e., the product of the diffusion coefficient (corresponding to the limit $q\rightarrow 0$) and $\tau_{\alpha}(q^*)$ is not independent of $T$.

This decoupling has been interpreted as a signature of increasingly heterogeneous dynamics in the liquid near its glass transition \cite{KumarEtal:JCP2006,sokolov09,EdigerDHReview_2000,GlotzerDHReview_2000}).  The term ``heterogeneous dynamics'' means that a glass former near $\Tg$ contains subensembles of particles with enhanced or reduced mobility relative to the average. To reveal this dynamic heterogeneity various methods were deployed \cite{EdigerDHReview_2000,GlotzerDHReview_2000}, including filtering techniques to track slow or fast particles, analysis of ensemble-averaged three- or four-point correlation functions \cite{BerthierEtal:JCP2007_1,BerthierEtal:JCP2007_2}, or scrutiny of the self-part of the van Hove correlation function \cite{FlennerSzamel:PRE2005_1,FlennerSzamel:PRE2005_2,ChaudhuriEtal:PRL2007}.

\begin{figure}
\includegraphics*[width=\colsize]{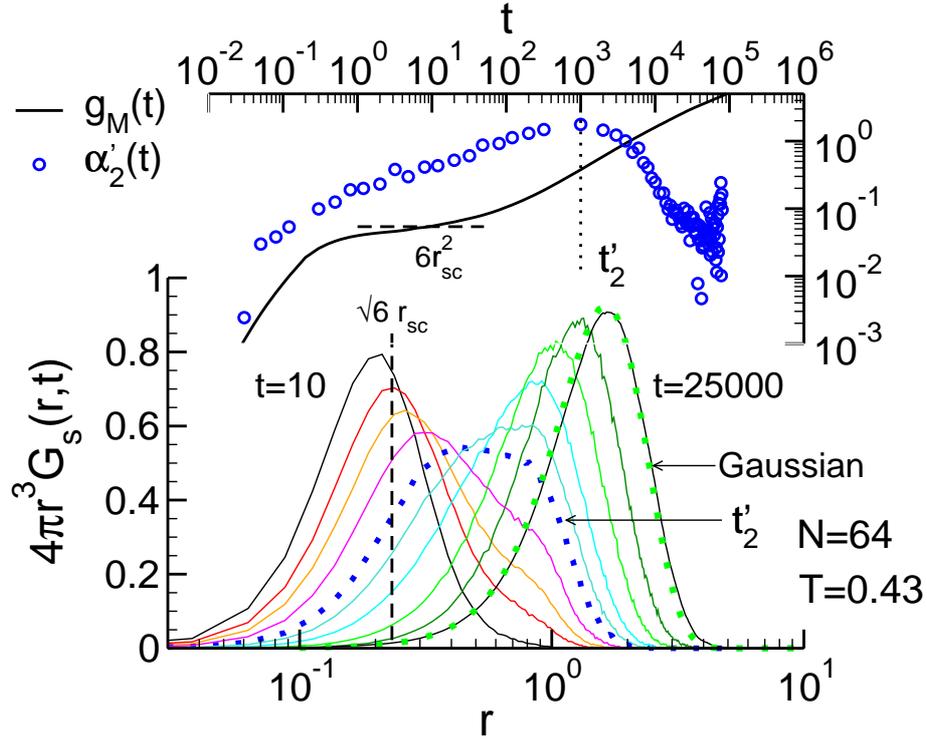}
\caption[]{Simulation results for a flexible bead-spring model: Probability distribution $P(\ln r;t)$ of the monomer displacements (left scale) at $T=0.43$ ($\Tc \simeq 0.415$) for $t=10$, 100, 250, 500, 1000, 1500, 2500, 5000, 10000, and 25000 (from left to right). The thick dotted line shows $P(\ln r;t)$ at $t=t_2'$, the peak time of $\alpha_2'(t)$. The dotted curve for $t=25000$ represents the Gaussian approximation for $P(\ln r;t)$.  The vertical dashed line indicates the displacement corresponding to the Lindemann localization length $\sqrt{6}r_{\mathrm{sc}}=0.232$. Right scale: Monomer MSD $g_\mathrm{M}(t)$ and non-Gaussian parameter $\alpha_2'(t)$ at $T=0.43$. The vertical dotted line indicates the peak time $t_2'$ of $\alpha_2'(t)$ and the horizontal dashed line the ``plateau value'' $6r_{\mathrm{sc}}^2$. Figure adapted from \cite{peter09}.}
\label{ngd_pp}
\end{figure}

Here we briefly discuss the approach of Refs.~\cite{FlennerSzamel:PRE2005_1,FlennerSzamel:PRE2005_2} which has recently been applied to simulation data of a flexible bead-spring model \cite{peter09}. Following \cite{FlennerSzamel:PRE2005_1,FlennerSzamel:PRE2005_2} we define the probability distribution $P(\ln r;t)$ of the logarithm of monomer displacements in time $t$ by
\begin{equation}
P(\ln r;t) = 4 \pi r^3 G_\mathrm{s}(r,t)\;,
\label{eq:defPlogr}
\end{equation}
where $G_\mathrm{s}(r,t)$ is the self-part of the van Hove function, as before. This probability is shown in \fref{ngd_pp}, together with $g_\mathrm{M}(t)$ and the non-Gaussian parameter \cite{FlennerSzamel:PRE2005_1,FlennerSzamel:PRE2005_2}
\begin{equation}
\alpha'_{2}(t)= \frac{g_{\mathrm{M}}(t)}{3} \,
\bigg \langle \frac{1}{|\vec{r}(t)-\vec{r}(0)|^2} \bigg \rangle -1 \; ,
\label{eq:a21}
\end{equation}
where $\vec{r}(t)$ denotes the position of a monomer at time $t$. In the cold melt there are clear deviations from Gaussian behaviour at all but the shortest and longest times.  The non-Gaussian parameter is positive and has a maximum at time $t_2'$ in the late-$\alpha$ regime.  At this time the distribution $P(\ln r;t)$ is very broad, exhibiting small ($r \gtrsim \sqrt{6}r_{\mathrm{sc}}$) and large ($r \gtrsim 1$) displacements. This hints at large, non-Gaussian fluctuations in particle mobility when $\alpha'_{2}$ peaks. On cooling toward $\Tc$ the broad displacement distribution develops into a double-peak structure, indicative of large disparities in particle mobility. Apparently, two populations of monomers coexist, ``slow'' ones which have not moved much farther than 10\% of their diameter in time $t_{2}'$, and ``fast'' ones which have left their cage and covered a distance of about their diameter or more.  This bimodal character of the structural relaxation is hard to predict from MCT \cite{FlennerSzamel:PRE2005_2}, could be responsible for the absence of the divergence of the relaxation time at $\Tc$, and appears to be a general feature of materials close to glass or jamming transitions \cite{ChaudhuriEtal:PRL2007}.

\subsection{Effects of confinement on glassy polymers}

\subsubsection{Brief overview of some experimental results}
During the past fifteen years the impact of geometric confinement on the glass transition has received considerable attention. The progress in the field is described in several comprehensive reviews \cite{ForrestDalnoki:AdvColSci2001,HartmannEtal:BroadbandDS2003,KremerEtal:BroadbandDS2003,RothDutcher:2004,AlcoutlabiMcKenna:JPCM2005,ChristianeEtal:JPCM2006,Tsui:Book2008}. The picture emerging from these studies is that glass formers confined to nanoscopic dimensions may exhibit deviations from bulk behaviour due to the interplay of spatial restrictions and interfacial effects. The latter (nonuniversal) effects can result from particle-substrate interactions, confinement-induced changes of the liquid structure or polymer conformations, density variations, etc.\ \cite{Donth,AlcoutlabiMcKenna:JPCM2005}, and often appear to dominate the behaviour of the confined glass former \cite{ChristianeEtal:JPCM2006}.

These interfacial effects have an important impact on the glass transition of thin polymer films. For films supported on a substrate, many studies \cite{KeddieEtal:EPL1994,SharpForrest:PRE2003,Forrest:EPJE2002,HerminghausEtal:EPJE2001,GrohensEtal:EPJE2002,FukaoEtal:PRE2000,FukaoEtal:PRE2001,KimEtal:Langmuir2001,TsuiZhang:Macro2001,fryer01,EllisonTorkelson:NatureMaterials2003,EllisonEtal:Macro2005,PriestleyEtal:PRE2007,LupascuEtal:MJNCS2006,NapolitanoWuebbenhorst:JPCB2007}, though not all \cite{EfremovEtal:Macro2004,HuthEtal:EPJS2007}, find reductions in $\Tg$ with decreasing film thickness ($\film$) if the polymer-substrate attraction is weak.  This is also the case for one of the most extensively studied systems, polystyrene on a variety of substrates \cite{KeddieEtal:EPL1994,Forrest:EPJE2002,HerminghausEtal:EPJE2001,FukaoEtal:PRE2000,FukaoEtal:PRE2001,KimEtal:Langmuir2001,TsuiZhang:Macro2001,EllisonTorkelson:NatureMaterials2003,EllisonEtal:Macro2005,PriestleyEtal:PRE2007,LupascuEtal:MJNCS2006,NapolitanoWuebbenhorst:JPCB2007}.  Here mesurements of surface relaxation after nanodeformation \cite{FakhraaiForrest:Science2008} or of the positional dependence of $\Tg$ \cite{EllisonTorkelson:NatureMaterials2003} further indicate that the depression of $\Tg$ is related to a ``free-surface effect'':  Monomers near the free surface are expected to be more mobile because they feel less steric contraints than their peers in the bulk. This enhanced mobility should lead to reductions in $\Tg$.

Measurements of (the average) $\Tg$ provide information on the dynamic reponse of the polymer films on the time scale associated with $\Tg$ which depends on experimental conditions, such as the cooling rate \cite{FakhraaiForrest:PRL2005}. Additionally, the full structural relaxation has also been explored by several techniques \cite{RothDutcher:2004,AlcoutlabiMcKenna:JPCM2005}, in particular by dielectric spectroscopy \cite{HartmannEtal:BroadbandDS2003,SharpForrest:PRE2003,FukaoEtal:PRE2000,FukaoEtal:PRE2001,Fukao:EPJE2003,PriestleyEtal:PRE2007,LupascuEtal:MJNCS2006,NapolitanoWuebbenhorst:JPCB2007,NapolitanoEtal:Macro2008,LundEtal:Macro2008,SergheiEtal:Macro2006,SergheiEtal:JNCS2007}.  Dielectric spectroscopy allows for a simultaneous measurement of $\Tg$ and the relaxation spectrum, even for thin films. A key finding of these studies is that the $\alpha$ process is broadened relative to the bulk. It is possible to interpret this broadening as a consequence of spatially heterogeneous dynamics in nanoconfinement.  At the interfaces the segmental dynamics can be enhanced (e.g.\ at the free surface) or slowed down (e.g.\ at an attractive substrate) relative to the bulk.  Chain segments in layers adjoining these interfacial layers should also have their dynamics perturbed, albeit to a lesser extent, which will lead to a still weaker perturbation for the next layer, and so on.  This interpretation implies that there is a smooth transition from interface-induced perturbations of the dynamics to bulk behaviour with increasing distance from the interfaces.
Recently, this idea has been exploited to analyze dielectric spectra of nanostructured diblock copolymer melts \cite{LundEtal:Macro2008}, and it is also consistent with the positional dependence of $\Tg$ found in \cite{EllisonTorkelson:NatureMaterials2003}.

\subsubsection{Simulation work on geometrically confined glass-forming polymer systems}
\label{subsubsec:confine}
Simulation studies of confined glass formers support this view of an interface-induced gradient in relaxation \cite{baschnagel05}. Many of these studies utilize simple models, e.g.\ binary liquids \cite{scheidler02,scheidler03,scheidler04,GalloEtal:EPL2002} or bead-spring polymer models \cite{torres00,jainPRL04,jain02a,bohme02,yoshimoto05,riggleman06,BaljonEtal:Macro2005,BaljonEtal:PRL2004,MoritaEtal:Macro2006,varnik02e,varnik02d,varnik02b,varnik03,peter06,peter07,peter08,peter09a,peter09b} (see however \cite{HudzinskyyEtal:2009,AlexiadisEtal:Macro2008,XuMattice:Macro2003,mansfield91,ManiasEtal:ColSurf2001} for work on chemically realistic models), to explore the impact of confinement in thin films \cite{scheidler02,scheidler04,torres00,jainPRL04,jain02a,bohme02,yoshimoto05,riggleman06,BaljonEtal:Macro2005,BaljonEtal:PRL2004,MoritaEtal:Macro2006,varnik02e,varnik02d,varnik02b,varnik03,peter06,peter07,peter08,peter09a,peter09b,XuMattice:Macro2003,mansfield91,ManiasEtal:ColSurf2001}, pores \cite{scheidler03} or systems containing nanofillers \cite{starr01,starr02,GalloEtal:EPL2002}. The simulations reveal a complex relaxation behaviour on approach to the glass transition of the confined glass former and also report shifts of $\Tg$, qualitatively similar to the trends sketched above for experiments. In the following we illustrate these results by some examples.

\begin{figure}
\includegraphics*[width=\colsize]{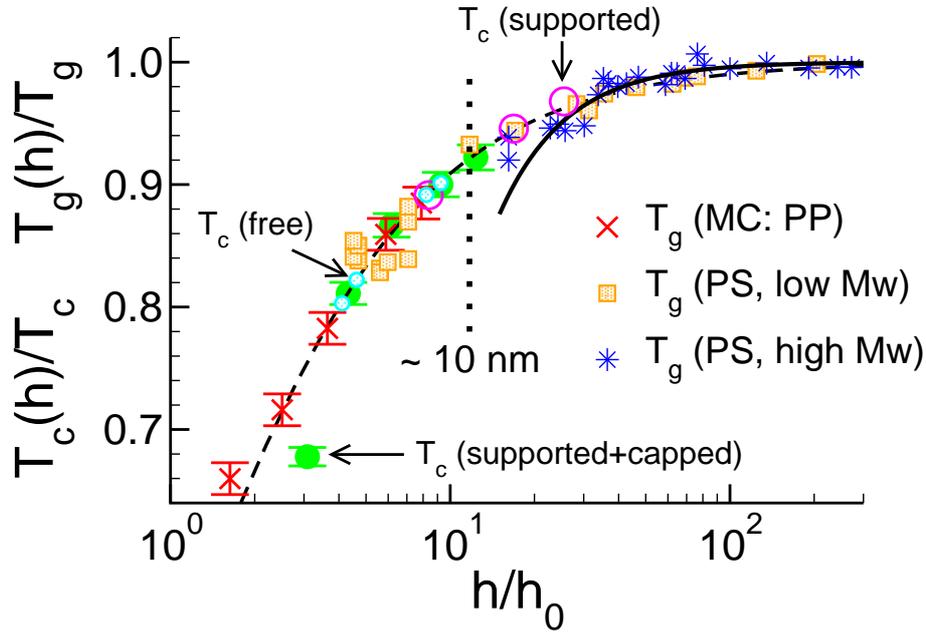}
\caption[]{$\Tc(h)/\Tc$ and $\Tg(h)/\Tg$ ($\Tc$ and $\Tg$ denote the bulk values) versus rescaled film thickness $h/h_0$. MD results for supported films (open circles), free-standing films (shaded circles), and films confined between two smooth repulsive walls (filled circles) are compared to the glass transition temperatures $\Tg(h)$ of three studies: (i) Monte Carlo simulations of a lattice model for free-standing atactic polypropylene (PP) films \cite{XuMattice:Macro2003} (crosses). (ii) Experiments of supported atactic PS films  of low molecular weight (open squares) \cite{HerminghausEtal:EPJE2001}. (iii) Experiments of supported, high-molecular weight PS films \cite{KeddieEtal:EPL1994} (stars).  The solid and dashed line show Eqs.~\eref{eq:TgKeddie} and \eref{eq:TgKim}, respectively. The vertical dotted line roughly indicates a film thickness of 10 nm. Figure adapted from \cite{peter07}.}
\label{fig:TcTg_vs_thickness_relative_values}
\end{figure}

\Fref{fig:TcTg_vs_thickness_relative_values} compares the reduction of $\Tg$ with film thickness found in experiments on supported PS films of low \cite{HerminghausEtal:EPJE2001} and high molecular weight \cite{KeddieEtal:EPL1994} with the $\Tg$ shifts obtained from simulations of free-standing films \cite{peter06,peter07,XuMattice:Macro2003}, supported films \cite{peter06,peter07}, and films confined between two substrates \cite{varnik02e,varnik02d}. The simulations study a chemically realistic model of polypropylene \cite{XuMattice:Macro2003} or flexible bead-spring models \cite{varnik02e,varnik02d,peter06,peter07}, and use as a substrate completely smooth walls which are either purely repulsive or weakly attractive.  Despite the obvious differences between experimental and computational systems---different polymers, absence or presence of a substrate, etc.---a master curve for the reduction of $\Tg$ can be constructed, if $\Tg(\film)$ is scaled by the bulk value and $h$ by a characteristic thickness $h_0$ that depends on the nature of the system, but only (very) weakly on molecular weight. This master curve allows us to compare the different systems, which is instructive in several respects. First, the close agreement of the PS data for low and high molecular weight suggests that possible changes of the entanglement density \cite{TsuiZhang:Macro2001,MeyerEtal:EPJ2006} or chain conformation \cite{CavalloEtal:JPCM2005,MeyerEtal:EPJ2006} in thin films are probably not responsible for the $\Tg$ depression. Furthermore, the experimentally observed $\Tg$ shifts have stimulated several attempts to model this phenomenon theoretically  \cite{HerminghausEtal:EPJE2001,Herminghaus:EPJE2002,PGG:EPJE2000,McCoyCurro:JCP2002,TruskettGanesan:JCP2003,MittalEtal:JPCB2004,Chow:JPCM2002,Ngai:EPJE2003,LongLequeux:EPJE2001,MerabiaEtal:EPJE2004,LipsonMilner:EPJB2009}. For instance, based on a percolation model for slowly relaxing domains Long and Lequeux \cite{LongLequeux:EPJE2001}
derive the formula (solid line in \fref{fig:TcTg_vs_thickness_relative_values})
\begin{equation}
\Tg(\film)=\Tg\bigg [1 -
\Big (\frac{\overline{\film}_0}{\film} \Big )^\delta \bigg] ,
\label{eq:TgKeddie}
\end{equation}
originally suggested by Keddie {\em et al} as an empirical parametrization in their seminal study on supported PS films \cite{KeddieEtal:EPL1994}. An alternative parameterization was proposed by Kim {\em et al}  \cite{KimEtal:Langmuir2001} (dashed line in \fref{fig:TcTg_vs_thickness_relative_values})
\begin{equation}
\Tg(\film)=\frac{\Tg}{1+\film_0/\film} ,
\label{eq:TgKim}
\end{equation}
and an attempt was undertaken to justify this formula by a viscoelastic capillary waves model \cite{HerminghausEtal:EPJE2001,Herminghaus:EPJE2002}. \Fref{fig:TcTg_vs_thickness_relative_values} shows both formulas and reveals that, for a typical range of experimental film thicknesses, $15 \lesssim \film/\film_0 \lesssim 300$, it is hard to decide whether \eref{eq:TgKeddie} or \eref{eq:TgKim} is more accurate. However, including the simulation data for $\film \lesssim 10$ nm, \eref{eq:TgKim} appears to provide the better description of the $\Tg$ shift. Therefore, \eref{eq:TgKim} will be employed later (\fref{fig:tau_from_g0_supported_layer_resolved_new_withTc}) in an analysis of the local $\Tg$ of simulated polymer films.

\begin{figure}
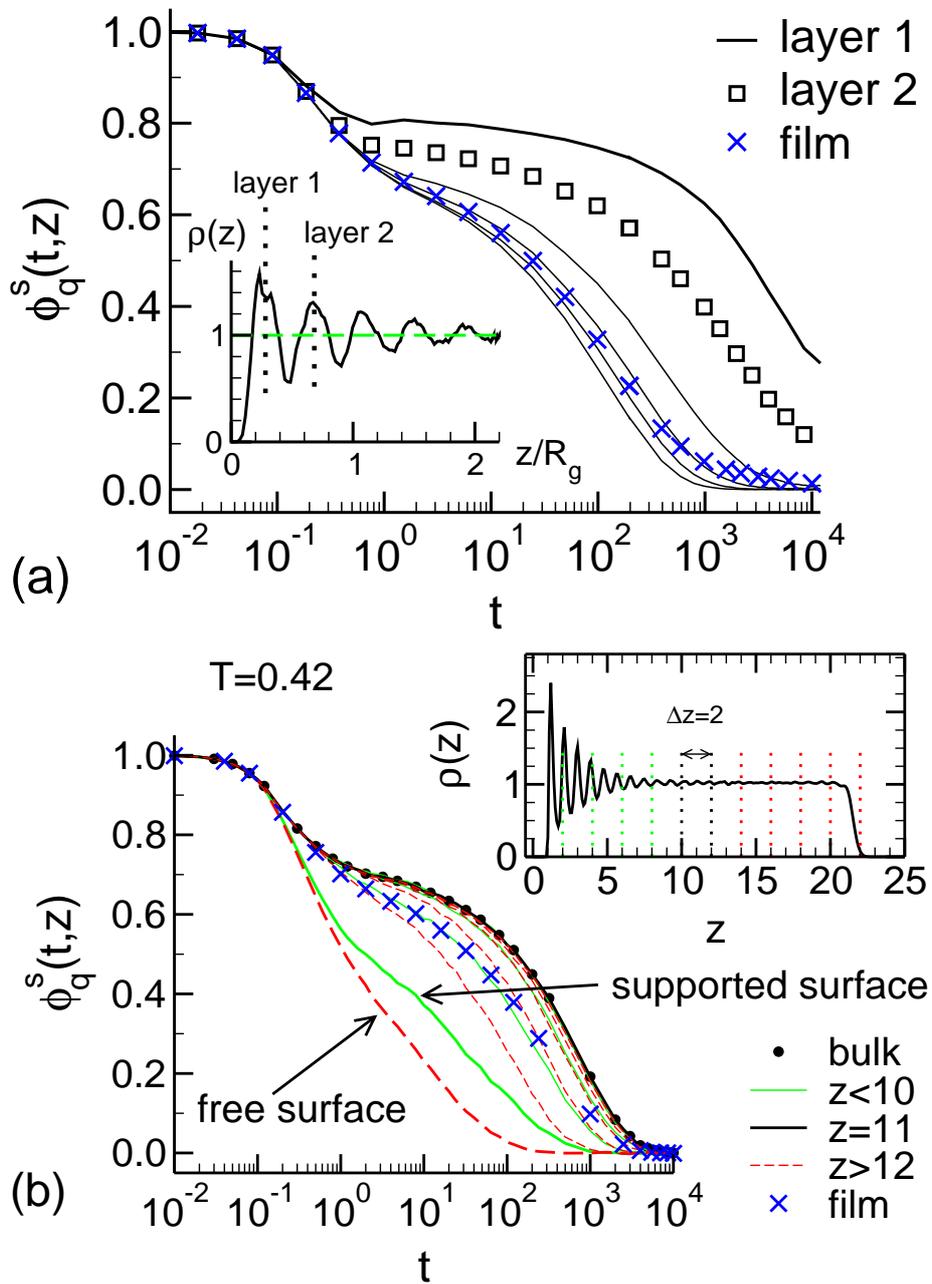

\includegraphics*[width=\colsize]{./icscf_attractiveFillers_FStarr_PRE2001}\\[2mm]
\includegraphics*[width=\colsize]{./incoherentscattering_layers}
\caption[]{(a) Layer-resolved incoherent scattering function $\phi_q^\mathrm{s}(t,z)$ at $T=0.4$ and $q=7.08$ (= maximum of $S(q)$) for a bead-spring polymer melt surrounding an icosahedral filler particle.  The chain length is $N=20$, the melt density is $\rho=1$ (dashed horizontal line in the inset), and the filler attracts the monomers more strongly than they attract each other in the bulk. The solid lines and the squares show $\phi_q^\mathrm{s}(t,z)$ for different distances $z$ from the surface of the filler particle. The location of the first two layers is illustrated in the inset which depicts the monomer density profile $\rho(z/\Rg)$ ($\Rg \simeq 2.17$).  The crosses indicate the average over all layers.  Figure adapted from \cite{baschnagel05}.
(b) $\phi_q^\mathrm{s}(t,z)$ at $T=0.42$  and $q=6.9$ ($\approx$ maximum of $S(q)$) for a bead-spring polymer melt in a supported film of thickness $h = 20.3$ ($\Tc \approx 0.392$). $z$ denotes the distance from the (left) wall.    $\phi_q^\mathrm{s}(t,z)$ is obtained as an average over all monomers which remain for all times shown in a layer of width $\Delta z = 2$ and centered at $z$.  The average behaviour of the film (average over all layers) is indicated by crosses and the bulk data by filled circles. Inset: Monomer density profile $\rho(z)$ versus $z$.  The layers for which $\phi_q^\mathrm{s}(t,z)$ is shown in the main figure, are indicated. Figure adapted from \cite{peter06}.}
\label{fig:incoherent_Peter_Starr}
\end{figure}

\begin{figure}
\includegraphics*[width=\colsize]{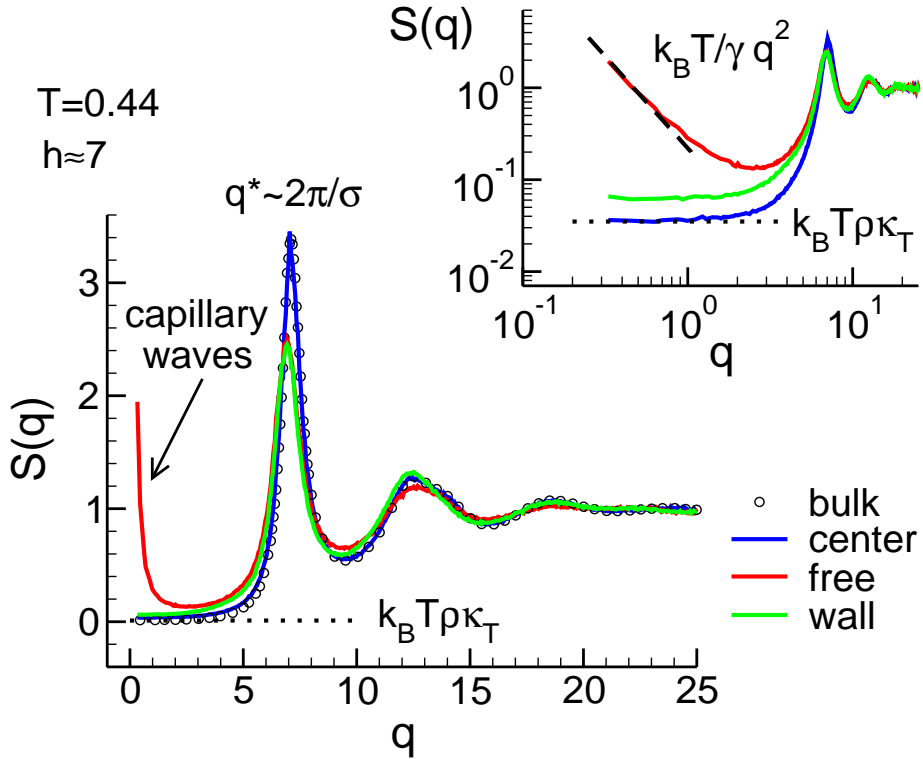}
\caption{Static structure factor $S(q)$ for a bead-spring melt ($N=10$) in the bulk (circles) and in supported film geometry (lines) at $T=0.44$ ($\Tc \approx 0.405$ in the bulk; $\Tc(\film \simeq 7) \approx 0.361$ in the film \cite{peter06}). For the film $S(q)$ is shown respectively for layers in film center, at wall and at the free surface. At the free surface the steep rise of $S(q)$ for small moduli $q$ of the wave vector is due to capillary waves. For the bulk the compressibility plateau $k_\mathrm{B}T \rho \kappa_T$ is indicated by a  horizontal dotted line ($\kappa_T$ denotes the isothermal compressibility). Inset: Same data as in the main figure, but in a log-log representation. The small-$q$ behaviour expected from capillary wave theory is depicted by a dashed line ($\gamma$ is the surface tension).}
\label{fig:Sq_supported_happrox7_T044_layers}
\end{figure}

The explanation for the $\Tg$ shift in the simulations rests upon the impact that the boundaries exert, because the average behaviour of the film---and so its $\Tg$---aggregates contributions from all layers in the film, and these layers have distinct properties. This can be illustrated by an analysis resolving structure and dynamics as a function of distance from the boundaries. For flexible bead-spring models \fref{fig:incoherent_Peter_Starr} shows two examples of such an analysis: the layer-resolved incoherent scattering function $\phi_q^\mathrm{s}(t,z)$ at $q^*$ (maximum of $S(q)$; cf.\ \fref{fig:Sq_supported_happrox7_T044_layers}) for a polymer melt surrounding a highly faceted, but nearly spherical filler particle with a structured and attractive surface (panel (a)) \cite{starr01}, and $\phi_{q^*}^\mathrm{s}(t,z)$ for a polymer film supported by a smooth, weakly attractive substrate (panel (b)) \cite{peter06}.

For the polymer melt surrounding the nanofiller the scattering function, averaged over all monomers (crosses in \fref{fig:incoherent_Peter_Starr}(a)), displays features unfamiliar from the bulk. The $\alpha$-relaxation seems to occur in two steps, as if there were two distinct processes, a fast one corresponding to a bulk-like phase far away from the filler and a slow one associated with interfacial relaxation. The layer-resolved analysis reveals that this interpretation is misleading. The strongly stretched tail of the average correlator results from the smooth gradient in the decay of $\phi_q^\mathrm{s}(t,z)$ which slows down on approach to the filler particle. References~\cite{starr01,starr02} show that the amplitude of this tail can be tuned by the monomer-filler interaction. Strong attraction leads to a more pronounced tail; vanishing attraction suppresses the tail. In the latter case, the shape of the scattering function is bulk-like.

Additional insight into the slow relaxation of particles in contact with a structured wall was obtained from studies of the self-part of the van Hove function \cite{scheidler04,SmithEtal:PRL2003,TeboulSimionesco:JPCM2002}. For times in the $\alpha$ regime and $T \gg \Tc$, $G_\mathrm{s}(r,t)$ has a clear two-peak structure. The first peak reflects particles that remain trapped in their cages, while the second peak, located on a length scale corresponding to the wall structure, reveals that particles have migrated to neighbouring wells.  Evidence for this kind of ``hopping motion'' is found for a binary LJ mixture \cite{scheidler04}, for a polymer melt adsorbed on a structured surface \cite{SmithEtal:PRL2003} or for model of liquid toluene confined in cylindrical mesopores \cite{TeboulSimionesco:JPCM2002}.  This relaxation mechanism could be generic when a liquid may lock into registry with the surface topography, leading to a mechanism of structural slowing down which coexists with glassy arrest at low $T$ \cite{Krakoviack:PRE2007,Krakoviack:PRE2009}

A similar locking is not possible at the free surface or a smooth, repulsive or weakly attractive wall. In thin films, one therefore expects enhanced dynamics relative to the bulk. \Fref{fig:incoherent_Peter_Starr}(b) shows that this expectation is borne out for flexible bead-spring models. For these models the free and smooth interfaces also create enviroments for nearby monomers which tend to reduce the cage effect: Local spatial correlations on the scale $q^*$ of the maximum of $S(q)$ are weaker in the films than in the bulk at the same temperature (\fref{fig:Sq_supported_happrox7_T044_layers}). This is an important contributing factor to the enhanced monomer dynamics found at both interfaces.

\begin{figure}
\includegraphics*[width=\colsize]{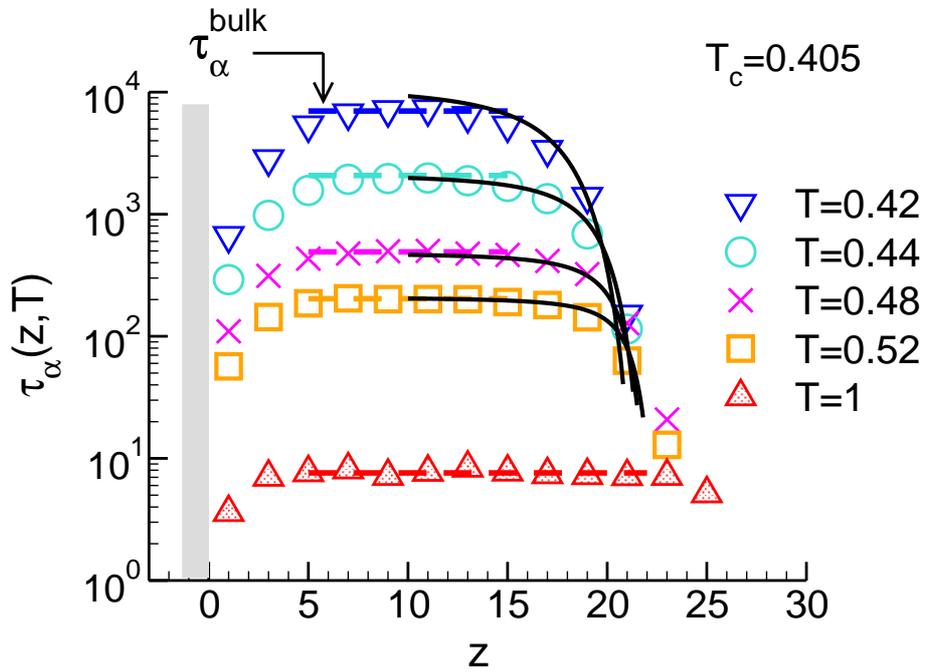}
\caption[]{Temperature dependence of the layer-resolved $\alpha$ relaxation time $\tau_\alpha(z,T)$ in a supported polymer film ($N=10$; $\Tc \approx 0.405$ in the bulk). The simulated system is the same as in \fref{fig:incoherent_Peter_Starr}(b). The solid lines at the free surface show \eref{eq:tauz} (no adjustable parameter). The horizontal dashed lines indicate the bulk values for $\tau_\alpha$ at the respective $T$. Figure adapted from \cite{peter07}.}
\label{fig:tau_from_g0_supported_layer_resolved_new_withTc}
\end{figure}

\Fref{fig:incoherent_Peter_Starr}(b) also shows that the interface-induced deviations from bulk dynamics continuously turn into bulk-like relaxation with increasing distance from the boundaries. The range of this crossover grows on cooling. \Fref{fig:tau_from_g0_supported_layer_resolved_new_withTc} illustrates this point by an analysis of the temperature dependence of the layer-resolved $\alpha$ relaxation time, $\tau_\alpha(z,T)$. At high temperature $\tau_\alpha(z,T)$ slightly deviates from the bulk value near the interfaces. With decreasing $T$ the interface-induced enhancement of the dynamics increasingly penetrates into the film and eventually propagates across the entire system for sufficiently low $T$.  This spatial dependence of $\tau_\alpha(z,T)$ can be modeled if two assumptions are made: (i) The average $\Tc(\film)$ of the film is given by \eref{eq:TgKim} and can be written as a ``democratic average'' \cite{LipsonMilner:EPJB2009} of the local $\Tc(z)$. That is,
\begin{equation}
\Tc(h) = \frac{\Tc}{1+\film_0/\film} = \frac 2h \int_0^{h/2} \mathrm{d}z \, \Tc(z) \;,
\label{eq:Tz1}
\end{equation}
which gives
\begin{equation}
\Tc(z) = \frac{\Tc (1 + h_0/z)}{(1 + h_0/2z)^2} \;.
\label{eq:Tz2}
\end{equation}
(ii) The second assumption is that the sole effect of the interface is to shift $\Tc$ from the bulk value to $\Tc(z)$, whereas all other parameters determining the $T$ dependence of $\tau_\alpha$ remain the same as in the bulk. Here we model this $T$ dependence by the MCT power law
\begin{equation}
\tau_\alpha(z,T) = \frac{\tau_0^\mathrm{bulk}}{(T - \Tc(z))^{\gamma_\mathrm{bulk}}} \; .
\label{eq:tauz}
\end{equation}
\Fref{fig:tau_from_g0_supported_layer_resolved_new_withTc} demonstrates that \eref{eq:tauz} yields a reasonable description of the increase of $\tau_\alpha$ from the free surface to the center of the film (solid lines in the figure) \cite{peter07}. The decrease of the $\alpha$ relaxation time on approach to the free surface therefore corresponds to a decrease of the local glass transition temperature, in qualitative agreement with the experimental results of \cite{EllisonTorkelson:NatureMaterials2003}

The propagation of enhanced or reduced mobility from the boundary toward the interior of the film has also been observed in other simulations on freely-standing \cite{jainPRL04} and supported polymer films \cite{BaljonEtal:PRL2004}.  These studies carried out a cluster analysis, of highly mobile monomers in the case of the freely-standing film and of immobile monomers for the supported film.  In both cases, it was found that clusters start at the interface and penetrate into the film.

In summary, simulation studies suggest that confined (polymeric) liquids display complex relaxation behaviour on approach to the glass transition because of the interplay of bulk-like slowing down of the dynamics and interfacial effects. Interfaces can enhance or retard the relaxation relative to the bulk. Enhanced relaxation may be expected for smooth or free interfaces, whereas strong particle-substrate attraction or particle caging in cavities of the substrate tend to slow down the dynamics. These interface-induced perturbations smoothly transition from the boundaries to the interior of the confined liquid. The range of this gradient grows on cooling so that the perburbations can propagate across the entire liquid for sufficiently strong confinement or low $T$. Similar interfacial effects may also be important for the analysis of other problems, for instance, for solvent evaporation from (spincoated) polymer films \cite{peter09a,peter09b} or for the modeling of the hydrodynamic boundary conditions in microfluidic devices \cite{ServantieMuller:PRL2008}. Therefore, it appears that those theoretical approaches, which treat the interplay of boundary effects and glassy slowing-down of the dynamics on the same (microscopic) footing, are most likely to advance our understanding in this field. Folding in boundary effects is certainly a major challenge \cite{FuchsKroy:JPCM2002,Krakoviack:PRE2007,Krakoviack:PRE2009}.  Simulations of model systems---as those reported here---should be helpful for the development of such theories.

\section{Studies of the glassy state}
\label{sec:belowTg}

Molecular simulations are intrinsically limited in terms of time scales. Therefore studying the glassy state, in which
the relaxation time scales are by definition extremely large, could seem to be out of reach for this kind of numerical approach. Paradoxically, however, the
fact that the intrinsic time scales of the system are large brings the simulations very close to actual experiments. The important   fact is that, deep in the glassy state, segmental relaxation times \cite{Angell97} greatly exceed both the experimental and the simulation time scales. For both  laboratory and computer   experiments, a time window is probed where segmental polymer motions are frozen and hence non-equilibrium phenomena associated with the glassy state are observed. The main issue is then rather a matter of sample preparation, which is usually done with much faster quenching rates in computer experiments {(cf.\ Sect.~\ref{subsubsec:Tg})}. As
we will see below, it turns out that the mechanical behaviour is not strongly affected, at least at a qualitative level, and that simulations can therefore be used
to analyse the structure and mechanical response of the glassy state with some confidence.

\subsection{Small deformations}

We start our discussion by considering the small strain, elastic part of the mechanical response  of a polymer glass, which can
be described at the macroscopic scale by linear elasticity.
It is now well recognized that, in spite of their homogeneity in density, glassy materials are heterogeneous at the nanoscale in terms
of their elastic properties. This heterogeneity is reflected indirectly in some vibrational properties  (e.g., the Boson peak  \cite{Tanguy2002}), and is most
obviously revealed by simulation, that allows one to compute elastic constants and study elastic response at various scales \cite{tsamados09}. For polymer glasses, the
first study of that kind was  presented in refs   \cite{yoshimoto04,yoshimoto05b}, following earlier studies that indicated size dependent results for the elastic constants of
nanometric systems \cite{vanvorkum}.  This study, which uses a simple definition
of local elastic constants based on a local calculation of the usual fluctuation formulae \cite{yoshimoto05b,barrat09}, shows clearly that the
material is inhomogeneous at scales of the order of 5 to 10 monomeric sizes. Small regions displaying negative elastic constants and therefore
would be unstable if they were not surrounded by regions with large local moduli.  These aspects are not specific to polymer systems, but can be observed in simple molecular or metallic glass formers \cite{mayr09,tsamados09}, and using more sophisticated definitions of the local elastic moduli, which points to their universal character. However, they are particularly important
when the polymer glass is cast in the form of a nanostructure, as the mechanical failure of such structures will be strongly affected by the presence of elastic heterogeneities.

While the response of a polymer glass for deformations smaller than a few percent can be described as elastic
from a macroscopic viewpoint, a detailed microscopic studies show that irreversible rearrangements take place at
low temperature even for very small deformations \cite{papakonstantopoulos08}. This irreversible behaviour is responsible for a dissipative part
or the mechanical response, sometimes described as anelasticity. Ultimately, this dissipation is associated with strongly localized plastic events that involve only a few monomers, and can be thought as the precursors of the dissipative processes that take place under larger deformation.

Two specificities of polymer glasses,already mentioned in Sect.~\ref{subsubsec:confine}, are the interest for thin films and the ability to introduce various types of additives, from small molecules to nanoparticle fillers. Relatively few studies are concerned with the thin film modifications of the glassy state itself. Jain and de Pablo \cite{jain04} showed that the vibrational density of states is
modified compared to the same glass in the bulk. In particular, the usual excess of low frequency modes observed in glasses (the so called boson peak) is enhanced in free standing thin films. This increase of the soft modes, low frequency part of the spectrum is consistent with a negative shift of the glass transition temperature. As the strength of the boson peak is in general, connected to
a decreased fragility (in the sense of Angell's classification of glasses, see \cite{AngellEtal_JAP2000}, this points to a reduction in fragility with film thickness.

Many studies on the other hand have been devoted to the modification of the glassy state under the influence of nanometric
 filler particles \cite{starr01,starr02,Papakonstantopoulos07}. As discussed in Sect.~\ref{subsubsec:confine} (see \fref{fig:incoherent_Peter_Starr}), the vicinity of fillers was shown to modify the dynamics of the polymer,
 with a slowing down and a shift of the glass transition temperature in the case of attractive interactions.
 From the mechanical point of view, it was also shown that the presence of attractive fillers results in the
 presence of a layer with enhanced mechanical properties (higher local moduli) \cite{papakonstantopoulos05,Papakonstantopoulos07} and modifies the structure of the entanglement network \cite{riggleman09a}.
 The spatial extension of this layer is of the order of a few monomer diameters and does not depend on the particle
 radius as soon as the latter exceeds 2 to 3 monomer size. As a result the mechanical properties of the composite are
  enhanced, however for  a well dispersed  composite this enhancement remains moderate.
  Note that the existence of a glassy layer around the nanoparticles is expected to affect the properties of the nanocomposite mostly in the region of the glass transition, where the contrast between this layer and the rest of the polymer is
  maximal \cite{Berriot2003,Montes2003}. As for polymer films, the fragility (inferred from boson peak strength and position) appears to be decreased with respect to the pure polymer.

Further studies of weak deformation include the influence of an external strain on the microscopic
dynamics \cite{riggleman07a} (in fact the latter study also extends to large, uniaxial deformations) , and the evolution of the creep compliance upon aging \cite{warren07}. It is found
a finite strain rate accelerates the segmental dynamics -measured by the bond orientation relaxation time- , whether
the corresponding strain is positive (dilation) or negative (compression). This symmetry between compression and dilation is indicative of a stress induced modification of the potential energy landscape, with the applied stress lowering the local barriers
and accelerating the caging dynamics. Free volume considerations, on the other hand, would not explain such a symmetry
\cite{riggleman07a}.  Acceleration of the dynamics under load (sometimes described as "mechanical rejuvenation") is also observed in the aging study of ref; \cite{warren07}, in which the creep compliance $J(t,t_w)$ of a simulated polymer is also shown to be well accounted for by the classical description of Struik \cite{struik}, in which the creep compliance is described as
a scaling function $J(t_{eff}/t_w^{\mu})$ where $t_w$ is the aging time, and
$t_{eff} = \int_0^t \left(\frac{t_w}{t_w+t'}\right)^{\mu} dt' $.
This result again shows that, in spite of the wide difference in time scales, the phenomenology of glassy polymers is reproduced by simulation work. It should also be noticed \cite{lee2009} that the relation between segmental mobility and strain rate is also dependent on the general strain history, so that, in contrast to the usual assumptions of Eyring's theory, no simple
mechanical variable can relate to the local mobility. Recent theoretical approaches \cite{ChenEtal:JPCM2009},
based on nonlinear Langevin equation description of segmental dynamics,  go beyond the simple Eyring description and account for this aging behaviour.

Finally, a rich and promising domain for simulation is the study of the complex effects of small molecules
on the glass transition and on the glassy state.  Such small molecules   can be described as {solvent}, plasticizers or antiplasticizers,
and in general their presence leads to an acceleration of the dynamics and a decrease in the glass transition
More precisely, plasticization of the polymer means that
the solvent does not only decrease $T_g$, but also softens the
polymer glass by reducing the elastic moduli.  Besides
this normally encountered case, there are also systems in
which the solvent antiplasticizes the polymer. That is,
the solvent decreases $T_g$, but increases the elastic moduli
of  the  polymer  glass.  The  origin  of  this  antiplasticizing
effect has recently been studied by simulations \cite{riggleman07b,riggleman07c},
and appears to be due to a more efficient packing associated with solvent molecules smaller than the monomers.
Small molecule fit into the "holes" of the polymer melt, and increase the elastic stiffness. However, their mobility also facilitates segmental mobility.
Both for antiplasticizers and plasticizers,   a strongly heterogeneous dynamics of the solvent molecules has been reported
\cite{riggleman07c,peter09}. The corresponding dynamical correlation length is however decreased in the presence of antiplasticizres,
with a fragility of the system that is also decreased.

\subsection{Large deformations and strain hardening}

The mechanical properties of glassy polymers at  large deformations {were} first investigated in the pioneering work of R\"ottler and Robbins \cite{rottler02,rottler03a,rottler03b,rottler05}, and their results were subsequently reproduced in a  number of studies \cite{SchnellPhD,makke09}. Beyond the peak stress, whose actual amplitude depends on the strain rate in a logarithmic way, as in simple
glasses, the  stress strain behaviour depends on the type of sollicitation. In pure shear or in simulations of a tensile test under triaxial conditions, the plastic flow proceeds through a cavitation of cavities and subsequent formation of fibrils. The drawing of these fibrils is essentially at constant stress.  This phenomenon is discussed in detail by R\"ottler \cite{RottlerJPCM2009}  Under uniaxial, almost volume conserving conditions (Poisson ratio close to 1/2),  a marked strain hardening is observed, i.e. the plastic stress increases with strain with a dependency that is close to linear. The corresponding deformation, however, is very far from elastic, and essentially non recoverable.

\begin{figure}
\includegraphics[width=\colsize]{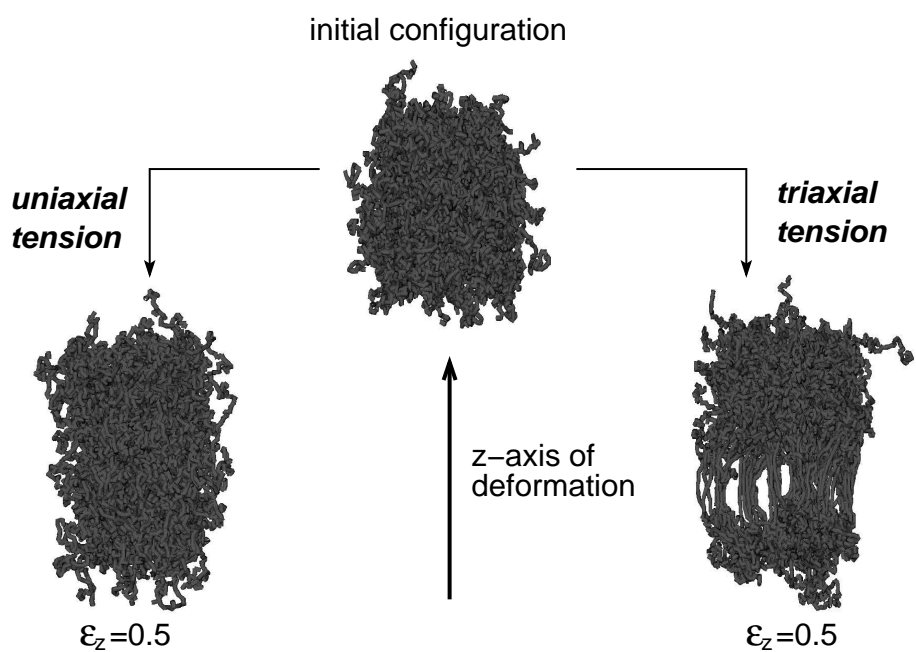}
\caption{Snapshots of three configurations of a polymer glass: initial undeformed state (up), uniaxially deformed configuration (left), and triaxially deformed configuration (right). Both deformed configurations correspond to a strain of 50\% along the $z$ axis of the simulation box. The simulation results are obtained from a flexible bead-spring model ($N=100$) at $T=0.2$ ($\Tg \approx 0.43$). Figure taken from \cite{SchnellPhD}.}
\label{fig:yieldcraz_art}
\end{figure}

  This peculiar property of polymer glasses, which strongly contributes to their practical applications, is the existence
  of the strain hardening regime at large deformation.
Large strain-hardening effect prevents the strain localization and leads to a tough response of the polymeric material; the material breaks only after a
significant plastic strain, as in the case for polycarbonate. For brittle polymers, as atactic
polystyrene, the strain-hardening effect is usually too weak, which leads to a brittle fracture due to crazes formation within a few percent of strain.  Understanding the microscopic origin of the strain-hardening effect provides a strategy for new tailor-made polymeric materials.

The microscopic  origin of this regime has been clarified only recently, thanks - to a large extent - to numerical simulations. Both the generic aspects of this phenomenon, and the way
strain-hard and strain-soft polymer materials differ in terms of segmental mobility and energetics, have been investigated
by extensive MD simulations.

Before we discuss simulation results, let us recall  the  popular rubber elasticity models \cite{Haward93}
which  predicts that the hardening modulus $G_h$ varies linearly with temperature $T$ and entanglement density $\rho_\mathrm{e}$, $G_h=\rho_\mathrm{e} k_\mathrm{B}T$. Experimental values, however, are two orders of magnitude larger, and, more important, show decrease with increasing  $T$ \cite{Kramer2005,Melick2003a}.
To take the dissipative nature of the plastic deformation into account the rubber elasticity model describes the stress-strain relation in the strain-hardening regime
with the help of two contributions \cite{Haward93}.

\begin{equation}
\sigma=\sigma_Y+c_2(\lambda^2-\lambda^{-1})
\label{eq:alexey1}
\end{equation}

The first part is a constant dissipative
stress $\sigma_Y$ , due to the presence of energy barriers. The
second is the strain-dependent part, which is thought to be
described by rubber-elasticity theory and represents the
strain-hardening effect. In this description the strain-hardening
modulus is not affected by thermally activated processes.

However, recent experiments demonstrate that such a description for the
strain-hardening part is invalid; the strain-hardening modulus
has characteristics of a thermally-activated process, and  decreases for
higher temperatures \cite{Govaert2004}. Secondly, at
higher strain rate the strain-hardening
modulus increases \cite{Sarva2007}, although for some polymers the dependency
on strain rate is rather weak \cite{Boyce90}. This increase can be
interpreted in a very elementary manner  within a barrier-crossing, Eyring like  picture, which however as discussed
above is known to be oversimplified.
Finally, the external pressure affects
the strain-hardening modulus as well. A higher external pressure
leads to an increase in the strain-hardening modulus \cite{vorselaars09a}. Again, this behaviour is typical for thermally-activated processes. All these three observations on the strain-hardening
modulus are not present within the classical rubber theory.
The failure of rubber-elasticity theory is due to the essential difference between a rubbery state, with many possible chain conformations between cross-links, and a non-ergodic glassy state where chain conformations are practically frozen, and transitions between different conformations are not possible.
Deformation of the polymer glass facilitates these transitions and is accompanied with the energy dissipation. The work for this irreversible energy dissipation is reflected in a
dissipative stress, which is absent in the rubber-elasticity theory. The dissipative nature of a polymer strain-hardening is confirmed by experiments. It is found that for PC and PS at large ($>15-30$\%) strains \cite{Salamatina88,Haward97,Oleinik2006}) more work is dissipated through heat than converted into internal energy.

The increase in stress for larger strains in combination with the
dissipative nature of the stress in the strain-hardening regime
suggest that there is an increase in the rate of energy dissipation,
i.e., more energy per unit of strain is needed for more stretched
samples to stretch them further. This picture is supported by the computer simulations of Hoy and Robbins \cite{Hoy2007a}, who  showed
that the
dissipative stress increases with larger strain and that at zero
temperature the stress was directly correlated to the rate of
changes in Lennard-Jones (LJ) binding.
Their more recent simulations of polymer toy models also demonstrated that most of the
stress at large strains is due to dissipation.
\cite{Hoy2007a,Hoy2008a}.

If the rubber elasticity theory is invalid,
what is then the polymer-specific part of the strain-hardening modulus?
Simulations of the mechanical deformation of simple molecular glasses
 show no strain hardening \cite{utz01}. Hence, for
short polymer chains the amount of strain hardening is expected to be small
as well. Moreover,  experiments \cite{Melick2003a} and simulations \cite{Hoy2006} show
that the strain hardening  modulus is positively correlated with the entanglement
density. Therefore one expects the strain hardening phenomenon to disappear progressively as the
chain length falls below the entanglement threshold.
Simulations show that this is indeed the case. In a
model polymer glass, Hoy and Robbins  \cite{Hoy2007a} demonstrated that there is a
gradual increase in the strain-hardening modulus as a function of
chain length for chains up to about the entanglement length. For
longer chains saturation in the modulus occurs. Similarly,
in MD simulations of atactic polystyrene \cite{Lyulin2005} only a weak  strain
hardening was observed for chain lengths of 80 monomers,
below the experimentally observed entanglement length of about
128--139 monomers \cite{Fetters94,Fetters99}.

 In the simulations of  ref. \cite{Hoy2007a}
it was also observed that the strain hardening is more correlated
with the change in the end-to-end distance of the polymer chains
than with the change in the global sample size.
If the sample size is decreasing while the end-to-end
distance does not, then the stress does not increase.

After the fundamental  aspects of strain hardening  have been clarified using simulations of simple, coarse-grained models,
more detailed studies using realistic models are required to understand the influence of chain architecture.
For example, it is known that
polymers with a larger persistence length often
have a higher strain-hardening modulus \cite{Haward93}.
It was even
shown by molecular-dynamics (MD) simulations that if the persistence
length of a polyethylene-like polymer is artificially increased by
changing the trans-to-gauche ratio, the strain-hardening modulus of
the resulting material increases as well \cite{Mckechnie93}. That
example illustrates that the strain-hardening modulus depends on the
conformation of the chain, which is frozen in the glassy state. The changes in this conformation
are associated with plastic events, that involve a collective nanoscale segmental dynamics of individual or neighbouring chains.  Understanding the differences in local mobility and dynamics  for chemically different polymers under large deformations, and how it relates to the different mechanical characteristics, is therefore an important question, which has been studied in relatively few cases.

\begin{figure}
\includegraphics*[width=\colsize]{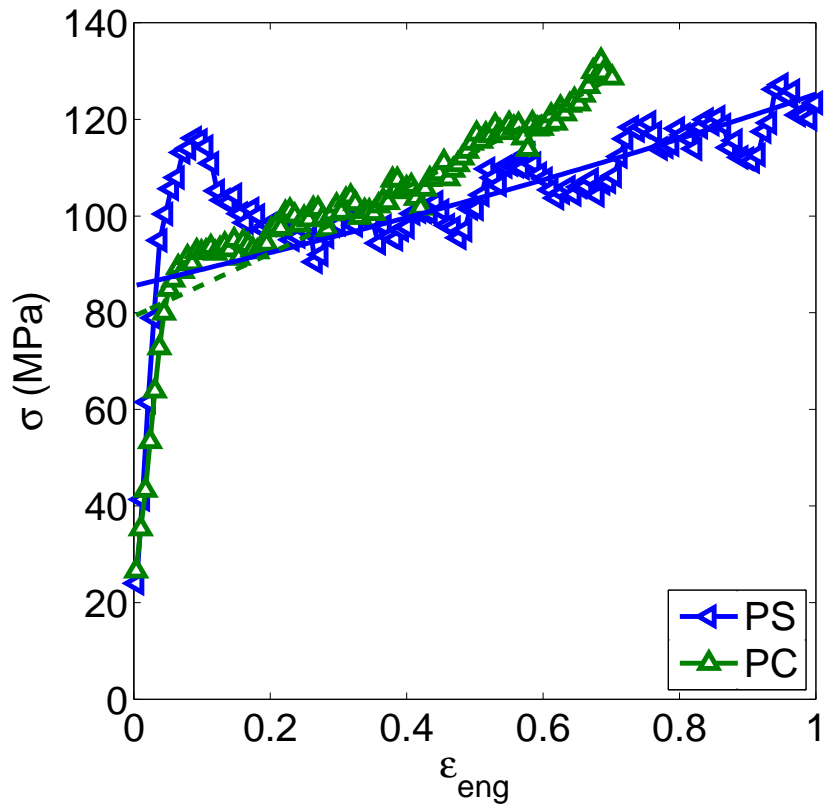}
\caption{The von Mises equivalent true stress $\sigma$ vs.\ strain
$\epseng$ for PS and PC. Solid lines are fits to
\eref{eq:alexey1}. Fit range is
$\epseng=0.3$--0.8. Note that the strain-hardening modulus $G_h$ for
PC (19 MPa) is almost twice that of PS (11 MPa), while their
extrapolated offset yield values $\sy$ are about the same (PC:
88 MPa; PS: 86 MPa). Adapted from ref. \cite{vorselaars09a}}
\label{fig:StressStrain-PSPC}
\end{figure}

As an example of such studies, and of the possibilities offered by simulation, we consider the extensive studies
 in refs. \cite{Lyulin2004,Lyulin2005,vorselaars09a,vorselaars09b} \r on  two glassy polymers that vary greatly in their strain-hardening moduli, viz.\ polystyrene (PS) and polycarbonate (PC, of which the modulus is more than a factor of two higher \cite{Melick2003b}). Molecular-dynamics simulations  \cite{Lyulin2004,Lyulin2005,vorselaars09a} have reproduced these experimental findings qualitatively, with a strain-hardening modulus of polystyrene that is much lower than that of polycarbonate, as shown in figure \ref{fig:StressStrain-PSPC}.
 They also allow one to identify the microscopic mechanisms responsible
for the observed difference in the strain-hardening modulus. In particular, they
show  the rate of non-affine displacements   (or local plastic events) increases with larger strain, and that the increase is
larger for PC. The non-affine displacements of particles are due to
restrictions and hindrances, in particular from covalent and steric
interactions. These restrictions are extremely important  at the scale of the covalent bond.
If particles would displace
affinely, then the equilibrium value of this chemical bond would be
excessively disturbed. To circumvent the bond stretch, the bond vector will not move affinely with the deformation.  At much larger length scales the situation is different. For a long chain the internal conformation can be adjusted, while still obeying to the equilibrium length of the covalent bond. The effective spring constant associated with the end-to-end distance is much less stiff. Moreover, due to the glassy state the relaxation time of spontaneous rearrangements at the scale of the whole chain greatly exceeds experimental and simulation time scales . Therefore, the end-to-end distance cannot adjust back towards the equilibrium value by means of spontaneous relaxations. Hence the end-to-end distance will follow the imposed deformation much more affinely. So we expect a more affine   response, as we probe larger length scales.

\begin{figure}
\includegraphics*[width=\colsize, angle=90]{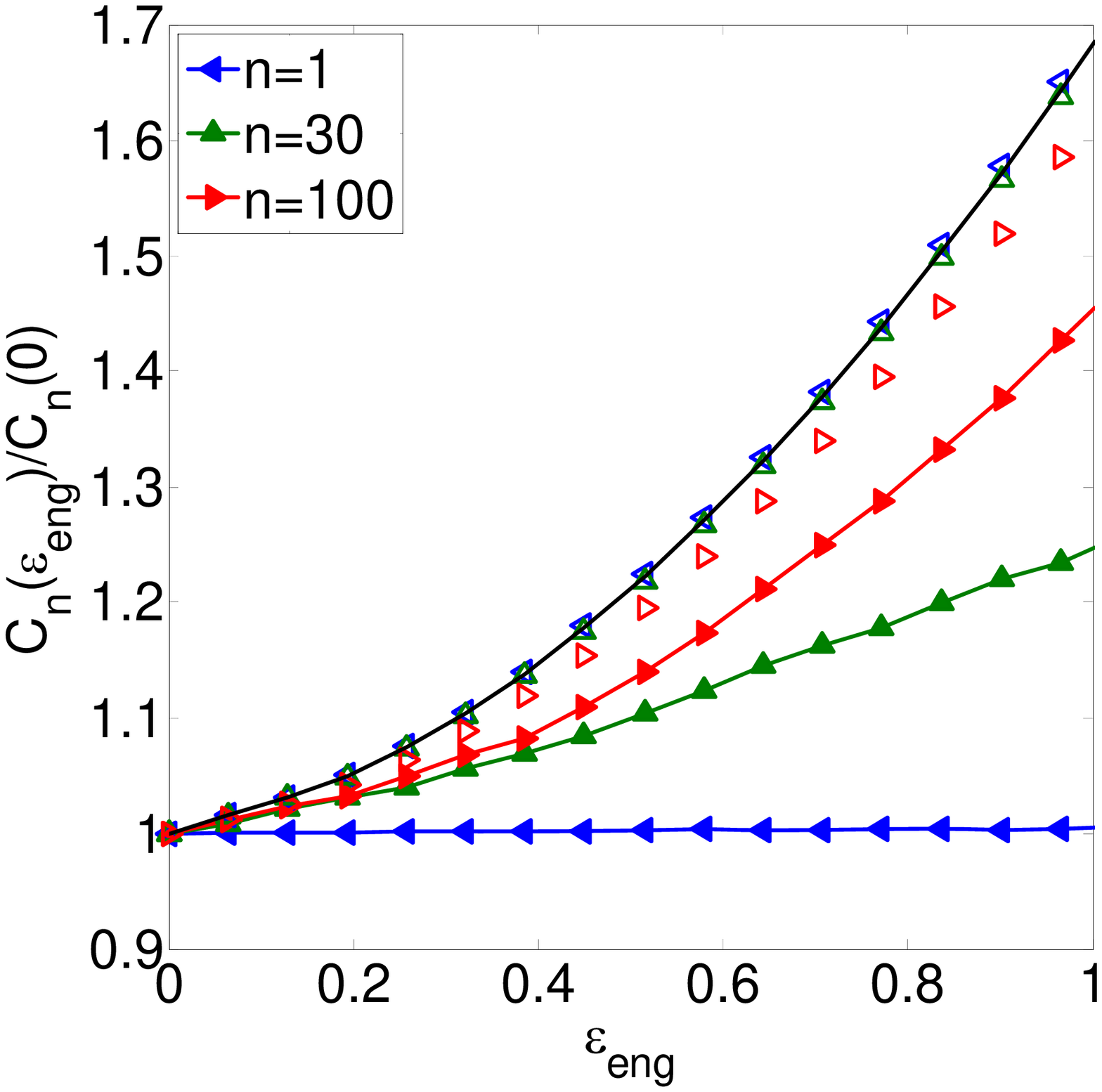}
\includegraphics*[width=\colsize, angle=90]{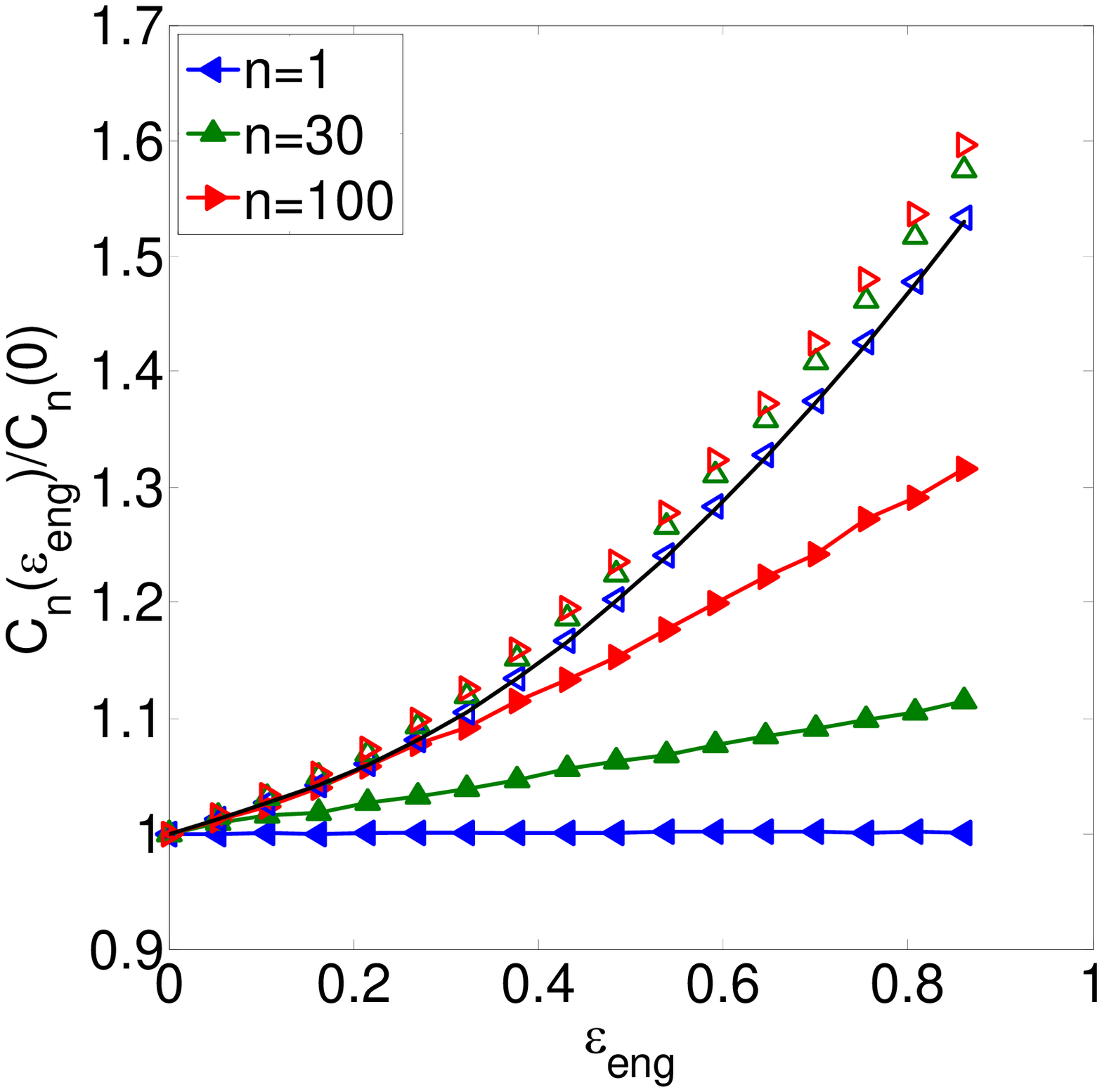}
\caption{The normalized characteristic ratio
$C_n'(\epseng)=C_n(\epseng)/C_n(0)$, which gives the deviation from affine
deformation as a function of strain, for a fixed {curvilinear distance $n$} along the chain backbone.  Upper panel:
PS. Lower panel:  PC.  Solid lines with filled symbols are simulation
results, while open symbols are results as if the sample would
deform in an affine way. The black curve is the affine
approximation $(\lambda_x^2+\lambda_y^2+\lambda_z^2)/3$. Adapted from ref. \cite{vorselaars09a}}
\label{fig:Norm-Cn-vs-eps-PSPC}
\end{figure}

This expectation is borne out by a detailed study of the affine character of the deformation of the polymer chain, as a function
of the internal distance along the chain \cite{vorselaars09b}.
At the scale of 100 chemical bonds, it is found that the deformation is essentially affine up to about 15\% strain for both PS and PC, while it is strongly non affine already at this strain at the scale of 30 bonds, see Fig. \ref{fig:Norm-Cn-vs-eps-PSPC}. For larger strains the relative deviation from affinity becomes larger for even long enough (100 backbone bonds) chains.  This effect is present in both PS and PC,
the difference being in  the
magnitude of the effect. For an internal distance of 30 monomeric units,
the deformation of the polycarbonate chain is only 30\% of the affine value at a strain of $50\%$,
while in polystyrene under the same conditions the chain deformation is 50\% of the affine value.
This
is because at the scale of the Kuhn length a chain cannot be
stretched any further.
As the Kuhn
length of polycarbonate is already larger than that of PS and the
total non-affine displacement of PC is to a large extent determined
by the backbone atoms, the increase in the effective stiffness
length leads to an increase in non-affine displacement, more energy
dissipation and hence a higher strain-hardening effect. For
polystyrene the Kuhn length is small and the major part of
non-affine displacement is not caused by the backbone. Hence the
expected increase in the effective stiffness length during
deformation does not lead to a substantial increase in plastic flow,
so that at moderate strains PS behaves more like a simple glass
without strain hardening, as opposed to polycarbonate.

\section{Perspectives}

The present review has attempted to describe some aspects of the recent progresses in our understanding of glassy
polymers that have been obtained from simulation work, with a particular focus on microscopic dynamic, glass transition  and mechanical properties. The results obtained in the past ten years show that the simulations have reached a state of maturity that allows them to address and clarify important issues in the field, in spite of the obvious  limitations in terms of length and time scales.
While the illustrative  examples  we have chosen  were relatively simple, they show that
the general phenomenology of the glass transition and the properties of the glassy state can be accounted for using
molecular simulations, in spite of the relatively small time scale that can be considered in such simulations.
We would like to close the discussion by suggesting some directions for future research, which
we believe will be soon - or are already - within the capabilities of state of the art simulations, and correspond to problems of practical or fundamental interest.

A major ingredient of the physics of glassy polymers - and more generally of glassy systems - is the so called
"time temperature superposition principle"  which assumes that relaxation processes can be  rescaled by using a single, temperature dependant relaxation time $\tau_\alpha(T)$. While this "principle" is widely used to produce frequency dependent relaxation function from data taken at various temperature, it is also well known to be applicable at best in the frequency region corresponding to the $\alpha$ peak, {presumably only above $\Tc$. It is a challenge to understand the molecular mechanisms underlying this superposition principle and the deviations thereof \cite{sokolov09}. A prominent feature of polymeric glasses
is the observation of secondary relaxations at higher frequency, the Johari-Goldstein $\beta$ processes (see Sect.~\ref{subsubsec:dynamics}). The associated relaxation times merge with the $\alpha$ relaxation times typically at temperatures of $0.8 T_g$, and the associated time scales are of the order of $10^{-6}$-$10^{-7}s$ at such temperatures.   Such time scales are beginning to be in the range
attainable by MD simulations, a fact that should allow one to clarify the molecular origins of secondary relaxations, their
relation to the existence of side groups,  to torsional modes, and other interpretations that have been put forward \cite{Mark}. Molecular simulation, by the flexibility it offers to block specific motions and to control chain properties
at various scales of coarse graining, should also allow one to investigate the putative relations between these secondary relaxations, fragility, and mechanical properties.

Nanocomposites and thin films have already been the subject of a number of simulation studies (see Sect.~\ref{subsubsec:confine}). Still, the
precise mechanisms that give rise to reinforcement and  nonlinear behaviour in nanocomposites, and to speed up of the dynamics in thin films, remain to be elucidated in detail.  In particular, for entangled polymers it can be expected
that the presence of interfaces modifies both the characteristics of the entanglement network and the monomeric friction coefficients. The interplay between these different aspects could, in principle, be elucidated in simulations of entangled systems. Again, the simulation times needed to explore such phenomena are extremely large, but, with the development of advanced simulation algorithms---as those alluded to in Sect.~\ref{sec:model}---and the increase of computer power they are starting to be within the attainable range.

Finally, we note that many polymer materials of practical interest display a mixed structure, being either partially crystalline
with an amorphous fraction, or chemically inhomogeneous with phases (or microphases in the case of block copolymers)
displaying different mechanical properties, e.g. glassy and rubbery.
The properties of the resulting nanocomposites have, up to now, received little attention from the standpoint of molecular simulation (see however \cite{ColmeneroArbe:SoftMatter2007,MorenoColmenero:JPCM2007,MorenoColmenero:PRL2008}). They are likely to be strongly influenced, especially for large deformations, by the interplay between chain architecture and material nanostructure. Another interesting phenomenon of that kind is strain induced crystallisation, which plays an important role in the strain hardening of some specific polymer. The capability of molecular simulation to describe polymer crystallization has been recently demonstrated \cite{MeyerMuellerPlathe:JCP2001,MeyerMuellerPlathe:Macromolecules2002,VettorelEtal:PRE2007}, so that the study of strain induced crystallisation  -as well as studies of semi crystalline phases appears to be an interesting goal for the future.

\paragraph{Acknowledgments}
The work reported in this article was carried out in fruitful collaboration with M. Aichele, N. Balabaev, C. Bennemann, K. Binder, S.-H. Chong, C. Donati, M. Fuchs, Y. Gebremichael, S. C. Glotzer, O. Lam\'e, A. Makke, M. Mareschal, H. Meyer, M.A.J. Michels, S. Napolitano, J. J. de Pablo, G. Papakonstantopoulos, W. Paul, M. Perez, S. Peter, R. Riggleman, B. Schnell, R. Seemann, F. W. Starr, F. Varnik, T. Vettorel, B. Vorselaars, and M. W\"ubbenhorst. It is a great pleasure to thank all of them.  We presented results from the research of W. Paul, G.~D. Smith, D. Bedrov and coworkers. We are grateful that they quickly provided the figures requested.  Our simulations were made possible by generous grants of computer time at the IDRIS in Orsay and at the NCF in Amsterdam,  JB gratefully acknowledges financial support by the IRTG ``Soft condensed matter''.

\bibliographystyle{unsrt}
\bibliography{all,additional_Joerg}

\end{document}